\documentclass[aps,floatfix,twocolumn,showpacs,prl,amsmath]{revtex4-1}
\usepackage{graphicx}
\usepackage{bm}
\usepackage{amssymb}
\usepackage{dcolumn}
\usepackage{subfigure}
\usepackage{threeparttable}
\usepackage{multirow}
\usepackage{mathrsfs}
\DeclareMathAlphabet{\mathscrbf}{OMS}{mdugm}{b}{n}
\usepackage[usenames,dvipsnames]{color}
\newcommand{\rmd}{{\rm d}}

\usepackage{mathtools}

\begin{document}
\newcommand{\vn}[1]{{\boldsymbol{#1}}}
\newcommand{\vht}[1]{{\boldsymbol{#1}}}
\newcommand{\matn}[1]{{\bf{#1}}}
\newcommand{\matnht}[1]{{\boldsymbol{#1}}}
\newcommand{\bege}{\begin{equation}}
\newcommand{\ee}{\end{equation}}
\newcommand{\bal}{\begin{aligned}}
\newcommand{\defbar}{\overline}
\newcommand{\SM}{\scriptstyle}
\newcommand{\eal}{\end{aligned}}
\newcommand{\udot}{\overset{.}{u}}
\newcommand{\exponential}[1]{{\exp(#1)}}
\newcommand{\phandot}[1]{\overset{\phantom{.}}{#1}}
\newcommand{\phandag}{\phantom{\dagger}}
\newcommand{\Trace}{\text{Tr}}
\newcommand{\Bxc}{\Omega}
\newcommand{\Torque}{\tau}
\newcommand{\gbar}{\bar{G}_{\vn{k} }^{\rm R}(\mathcal{E})}
\newcommand{\ebar}{\bar{\mathcal{E}}}
\newcommand{\vso}{\mathcal{V}_{\vn{k}}}
\newcommand{\intene}{\int\!\!\rmd\mathcal{E}}
\newcommand{\intkspa}{\int\!\!\frac{\rmd^d k}{(2\pi)^d}}
\setcounter{secnumdepth}{2}
\title{
The relation of the Dzyaloshinskii-Moriya interaction
to spin currents and to the spin-orbit field}
\author{Frank Freimuth}
\email[Corresp.~author:~]{f.freimuth@fz-juelich.de}
\author{Stefan Bl\"ugel}
\author{Yuriy Mokrousov}
\affiliation{Peter Gr\"unberg Institut and Institute for
Advanced Simulation,
Forschungszentrum J\"ulich and JARA, 52425 J\"ulich, Germany}
\date{\today}
\begin{abstract}
Starting from the general Berry phase theory of the
Dzyaloshinskii-Moriya interaction (DMI) we derive an expression
for the linear contribution
of the spin-orbit interaction (SOI).
Thereby, we show analytically that at the first order in SOI
DMI is given by the ground-state spin current.
We verify this finding numerically by \textit{ab-initio} calculations
in Mn/W(001) and Co/Pt(111) magnetic bilayers. 
We show that despite the strong SOI from the 5$d$ heavy metals
DMI is well-approximated by the first order in SOI, while the ground-state
spin current is not.
We decompose the SOI-linear contribution to DMI into two parts.
One part has a simple interpretation in terms of the Zeeman interaction
between the spin-orbit field and the spin misalignment
that electrons acquire in magnetically noncollinear textures. 
This interpretation provides also an intuitive
understanding of the symmetry of DMI on the basis of the spin-orbit field
and it explains in a simple way why DMI and ground-state spin currents
are related. 
Moreover, we show that energy currents driven by magnetization
dynamics and associated to DMI can be explained by counter-propagating
spin currents that carry energy due to their Zeeman interaction with
the spin-orbit field. Finally, we discuss options to modify DMI by
nonequilibrium spin currents excited by electric fields or light.
\end{abstract}

\pacs{72.25.Ba, 72.25.Mk, 71.70.Ej, 75.70.Tj}

\maketitle
\section{Introduction}
Most excitement about spin-currents arises from the prospects to
use them to transmit information 
dissipationlessly~\cite{dissipationless_spin_current},
to switch magnetic bits~\cite{current_induced_switching_using_spin_torque_from_spin_hall_buhrman,CoPtAlO_perpendicular_switching_Gambardella} 
and to move domain walls~\cite{chiral_domain_wall_motion_parkin,chiral_domain_wall_motion_beach}.
Therefore, in many cases the
generation of spin currents by applied electric fields, i.e., 
the spin Hall effect~\cite{rmp_she}, or by
magnetization dynamics, i.e., spin-pumping~\cite{enhanced_gilbert_thin_films},
or by laser excitation~\cite{thz_spin_current_kampfrath} are considered.
However, spin currents exist also in the absence of applied electric fields,
magnetization dynamics and laser pulses, when the system is in its ground state. 
These ground-state spin currents mediate important effects and interactions
as well. For example, in magnetic bilayer systems the ground-state spin
current flowing between the magnetic layer and the normal metal substate
when the magnetization is tilted away from the easy axis
provides the nonlocal contribution to the magnetic anisotropy torque~\cite{ibcsoit}.
Furthermore, the interlayer exchange coupling between magnetic layers 
in spin valves is mediated by ground-state spin 
currents~\cite{interlayer_exchange_coupling}. 

Recently, it has been proposed to estimate the Dzyaloshinskii-Moriya 
interaction (DMI) from the ground-state spin current~\cite{dmi_doppler_shift}. 
DMI describes the linear-in-$q$ contribution to the energy dispersion
$E(q)$ of spin-spirals and arises in systems with inversion asymmetry 
and spin-orbit interaction (SOI)~\cite{dmi_moriya,dmi_dzyalo}.
In several spintronics concepts DMI plays a central role.
Notably, it is a key ingredient to achieve current-driven
domain wall motion at high speed~\cite{chiral_domain_wall_motion_parkin,chiral_domain_wall_motion_beach} 
and to stabilize skyrmions~\cite{skyrmion_ground_states}.
The relation of DMI to the ground-state spin current also leads to
an intuitive interpretation of DMI as a Doppler shift~\cite{dmi_doppler_shift}.

Since the computational
evaluation of the ground-state spin current is easier than the
usually applied methodology for the calculation of the DMI 
coefficients from the $q$-linear contribution to the energy
dispersion $E(\vn{q})$ of spin
spirals~\cite{heide_dmi_few,heide_dmi_mnw,dmi_spirals_first_principles_heide,dmi_copt_thiaville},
it has been proposed to use the ground-state spin currents
for example to study the dependence of DMI on strain 
and voltage~\cite{dmi_doppler_shift}. 
However, the relation between DMI and ground-state spin current has
been derived in the strong exchange limit and
the accuracy of this spin-current approach to DMI has been demonstrated only in
the B20 compounds Mn$_{1-x}$Fe$_{x}$Ge 
and Fe$_{1-x}$Co$_{x}$Ge. 
In Co/Pt magnetic bilayer systems SOI
is much stronger than in these B20 compounds and
the applicability of the spin-current approach to such magnetic bilayer
systems has not been demonstrated yet.

The description of ferroelectric polarization by the Berry 
phase~\cite{polarizationtheory1},
the use of the Berry curvature in the theory of the anomalous Hall
effect~\cite{rmp_ahe},
and the discovery of topological insulators are well-known
examples for the success story of   
the quest for effects of topological and geometrical origin in 
band theory. 
Recently, it has been shown that the
exchange parameters can be expressed in terms of geometrical
properties such as the quantum metrics~\cite{geotexmosoi} and that
DMI can be expressed in
terms of the mixed Berry 
curvature ${\rm Im}\langle \partial u_{\vn{k}n}/\partial \hat{\vn{n}}|\partial u_{\vn{k}n}/\partial \vn{k}\rangle$, 
where derivatives with respect
to the $k$-point and derivatives with respect to the magnetization
direction $\hat{\vn{n}}$ are combined~\cite{mothedmisot,phase_space_berry,itsot}.

This Berry phase approach expresses DMI
directly in terms of the electronic structure, similar to the spin-current approach. 
This is a major advantage
of these two approaches compared to the spin-spiral method,
where DMI is extracted from  
the $q$-linear contribution to the energy
dispersion $E(\vn{q})$ of spin
spirals. While the relation of DMI to other important spintronics
effects is not directly obvious within the spin-spiral approach,
the Berry phase theory of DMI shows how
DMI is related to direct, inverse and thermal spin-orbit torques and to 
the twist-torque moments of wave-packets~\cite{phase_space_berry,mothedmisot,itsot}.
Since the spin-current approach to DMI establishes the connection to the
ground-state spin current, even more insights can be expected from
investigating the relationship between the Berry phase approach on the one hand
and the spin current approach on the other hand. 

In the present paper we use first-order perturbation theory to 
derive expressions for DMI and for the ground-state spin current,
which are valid at the first order in SOI. 
Thereby we find analytically that DMI is given
exactly by the ground-state spin current at the first order in SOI, which
clarifies the relation between the Berry-phase approach and the
spin-current approach to DMI.
Tuning the SOI strength in Co/Pt(111) and Mn/W(001)
magnetic bilayers artificially we confirm this analytical result numerically
by \textit{ab-initio} calculations. 
By studying both DMI and ground-state spin current as a function of SOI
strength we illustrate the limitations of the spin-current approach to DMI,
which breaks down for large SOI strength.
We find that the SOI-linear 
contribution to the ground-state spin current consists of two terms, 
and provide an intuitive interpretation of both of them.
We discover that one contribution is intimately linked to the
misalignment that conduction electron spins acquire as they traverse
magnetically noncollinear textures. This contribution 
can be understood as Zeeman interaction between
the spin-orbit field and the spin misalignment. 
When the magnetization direction is time-dependent, the spin misalignment
leads to counterpropagating spin currents, based on which
we elucidate the nature of the energy 
current 
that is driven by magnetization dynamics in systems with DMI~\cite{itsot}.
Thereby our theory exposes the connections of DMI to other spintronics concepts
such as spin-orbit fields and spin-transfer torque. 

This article is structured as follows.
In section~\ref{sec_first_order} we derive expressions for the SOI-linear 
contributions to DMI and to the ground-state
spin current and show that both agree within the first order perturbation theory in SOI.
In section~\ref{sec_gauge} and~\ref{sec_ajII} we interpret the two contributions to DMI
that arise at the first order in SOI.
In section~\ref{sec_interpretation_ground_state_energy_current} we explain how the
ground-state energy current associated with DMI can be understood from
counter-propagating spin currents driven by magnetization dynamics.
In section~\ref{sec_rashba_model} we show that ground-state spin currents exist for
the nonmagnetic Rashba model with zero DMI, but that these spin currents arise at the
third order of SOI.
In section~\ref{sec_ab_initio} we present \textit{ab-initio} calculations of DMI and
of ground-state spin currents in Mn/W(001) and Co/Pt(111) magnetic bilayers.
In section~\ref{sec_summary} we conclude with a summary and outlook, where we also
discuss the option to modify DMI by nonequilibrium spin currents excited by electric fields or light.
\section{First-order contribution of SOI to DMI}
\label{sec_first_order}
Due to DMI
the free energy density $F(\vn{r})$ at position $\vn{r}$ 
contains a term linear in the gradients of 
magnetization~\cite{mothedmisot}:
\bege\label{eq_first_order_free_energy}
F^{\rm DMI}(\vn{r})=
\sum_{j}
\vn{D}_{j}(\hat{\vn{n}}(\vn{r}))
\cdot
\left[
\hat{\vn{n}}(\vn{r})\times\frac{\partial \hat{\vn{n}}(\vn{r})}{\partial r_{j}}
\right],
\ee  
where $\vn{D}_{j}(\hat{\vn{n}})$ are the DMI coefficient vectors, which
depend on the magnetization direction $\hat{\vn{n}}(\vn{r})$ in 
systems where DMI is anisotropic.
The index $j$ runs over the three
cartesian directions $x$, $y$ and $z$.

Within the Berry phase
approach $\vn{D}_{j}(\hat{\vn{n}})$
is given by
\bege\label{eq_dmi_berry}
\vn{D}_{j}(\hat{\vn{n}})
\!=\!\!\!\int\!\!\!
\frac{\rmd ^d k}{(2\pi)^{d}}
\sum_{n}
\Bigl[
 f(\mathcal{E}_{\vn{k}n})\vn{A}_{\vn{k}nj}(\hat{\vn{n}})
-g(\mathcal{E}_{\vn{k}n})
\vn{B}_{\vn{k}nj}(\hat{\vn{n}})
\Bigr],
\ee
where $d$ is the dimension ($d$=2
or $d$=3), $f(\mathcal{E}_{\vn{k}n})=[1+e^{\beta(\mathcal{E}_{\vn{k}n}-\mu)}]^{-1}$ is
the occupation number of band $n$ at $k$ point $\vn{k}$,
$g(\mathcal{E}_{\vn{k}n})=-k_{\rm B}T\ln[1+e^{-\beta(\mathcal{E}_{\vn{k}n}-\mu)}]$ is the
contribution of the state $|u_{\vn{k}n}\rangle$ with band energy
$\mathcal{E}_{\vn{k}n}$ to the grand canonical potential, $T$ is
the temperature, $k_{\rm B}$ is Boltzmann's constant, $\mu$ is the
chemical potential and $\beta=(k_{\rm B}T)^{-1}$.
In the mixed Berry curvature~\cite{phase_space_berry}
\bege
\vn{B}_{\vn{k}nj}(\hat{\vn{n}})=
-2
\left[
\hat{\vn{n}}\times
{\rm Im}
\left\langle
\frac{\partial u_{\vn{k}n}}{ \partial\hat{\vn{n}} }
\left|
\frac{\partial u_{\vn{k}n}}{\partial k_{j}}\right.
\right\rangle
\right]
\ee
$\vn{k}$-derivatives are mixed with $\hat{\vn{n}}$-derivatives.
The twist-torque moment of wavepackets is
described by~\cite{mothedmisot}
\bege
\vn{A}_{\vn{k}nj}(\hat{\vn{n}})=
-
\left[
\hat{\vn{n}}\times
{\rm Im}
\left\langle
\frac{\partial u_{\vn{k}n}}{ \partial\hat{\vn{n}} }
\right|
\!\Bigl[
\mathcal{E}_{\vn{k}n}-H_{\vn{k}}
\Bigr]
\!\left|
\frac{\partial u_{\vn{k}n}}{\partial k_{j}}
\right\rangle
\right],
\ee
where $H_{\vn{k}}$ is the Hamiltonian in crystal-momentum representation,
i.e., $H_{\vn{k}}|u_{\vn{k}n}\rangle=\mathcal{E}_{\vn{k}n}|u_{\vn{k}n}\rangle$.

In the limit of strong exchange or small SOI the
DMI coefficient vector $\vn{D}_{j}$ 
can also be determined from the ground-state spin 
current $\vn{Q}_{j}$~\cite{dmi_katsnelson,dmi_doppler_shift}.
When the
ground-state spin current flowing in $j$ direction is defined as
\bege\label{eq_spincurr}
\vn{Q}_{j}=\frac{\hbar}{4}\intkspa
\sum_{n}f(\mathcal{E}_{\vn{k}n})
\left\langle
u_{\vn{k}n}
\right|
\{
\vn{\sigma},v_{j}
\}
\left|
u_{\vn{k}n}
\right\rangle,
\ee
the DMI coefficient vector can be written as 
\bege\label{eq_dmi_spincurr}
\vn{D}_{j}=-\vn{Q}_{j}.
\ee
Here, $v_{j}$ is the $j$ component of the velocity operator $\vn{v}$ 
and $\vht{\sigma}=(\sigma_x,\sigma_y,\sigma_z)^{\rm T}$
is the vector of Pauli
spin matrices. 

Both Eq.~\eqref{eq_dmi_berry} and
Eq.~\eqref{eq_spincurr} express the 
DMI directly in terms of the electronic structure,
which is a major advantage over the 
spin-spiral method, where DMI is extracted from
the $q$-linear term in the energy 
dispersion~\cite{heide_dmi_few,heide_dmi_mnw,dmi_spirals_first_principles_heide,dmi_copt_thiaville}.
However, Eq.~\eqref{eq_dmi_berry} and Eq.~\eqref{eq_spincurr} look very
different and the relationship between both approaches is not clear.
Since DMI arises at the first order in SOI~\cite{dmi_moriya},
we will use first-order perturbation theory to determine the
SOI-linear contributions to $\vn{D}_{j}$ and $\vn{Q}_{j}$.
This will facilitate the comparison between the approaches 
Eq.~\eqref{eq_dmi_berry} and Eq.~\eqref{eq_spincurr}
and elucidate their relationship.

Within Kohn-Sham density functional theory
interacting many-electron systems are described by the
effective single-particle Hamiltonian
\bege
\begin{aligned}
\label{eq_ks_hamil}
H(\vn{r})&=-\frac{\hbar^2}{2m}\Delta+V(\vn{r})+
\mu_{\rm B}\vn{\sigma}\cdot\hat{\vn{n}}\Bxc^{\rm xc}(\vn{r})+\\
&+
\frac{1}{2 e c^2}\mu_{\rm B}
\vn{\sigma}\cdot
\left[
\vn{\nabla}V(\vn{r})\times\vn{v}
\right]
,
\end{aligned}
\ee
where the first term describes the kinetic energy and
the second term is the scalar effective potential.
The third term describes the exchange interaction,
where 
$\mu_{\rm B}$ is the
Bohr magneton, $\Bxc^{\rm xc}(\vn{r})=\frac{1}{2\mu_{\rm
    B}}\left[V^{\rm eff}_{\rm minority}(\vn{r})-V^{\rm eff}_{\rm
    majority}(\vn{r}) \right]$
is the exchange field, $V^{\rm eff}_{\rm minority}(\vn{r})$ is the
effective potential of minority electrons and
$V^{\rm eff}_{\rm majority}(\vn{r})$ is the effective potential of
majority electrons. The last term is the spin-orbit interaction, 
where $e$ is the elementary positive charge and $c$ is the velocity of light.

The Hamiltonian in crystal-momentum representation is given
by $H_{\vn{k}}=e^{-i\vn{k}\cdot\vn{r}} H e^{i\vn{k}\cdot\vn{r}}$.
We decompose $H_{\vn{k}}$ into the spin-orbit
interaction $\mathcal{V}_{\vn{k}}$ and the Hamiltonian
$\bar{H}_{\vn{k}}$ without SOI, such that
\bege
H_{\vn{k}}=\bar{H}_{\vn{k}}+\mathcal{V}_{\vn{k}}.
\ee
We introduce the parameter $\xi$ to scale SOI up or down
and consider the modified Hamiltonian
\bege
H_{\vn{k}}^{\xi}=\bar{H}_{\vn{k}}+\xi\mathcal{V}_{\vn{k}}.
\ee
The DMI coefficient vector $\vn{D}_{j}$ of a system 
described by $H_{\vn{k}}^{\xi}$ can be
written as a power series with respect to $\xi$: 
\bege\label{eq_expand_dmi}
\vn{D}_{j}=\xi \vn{D}^{(1)}_{j}+\xi^2 \vn{D}^{(2)}_{j}+\xi^3 \vn{D}^{(3)}_{j}+\dots,
\ee
where $\vn{D}^{(1)}_{j}$ is linear in $\mathcal{V}_{\vn{k}}$.

In order to derive an explicit expression for $\vn{D}^{(1)}_{j}$ it 
is convenient to rewrite 
Eq.~\eqref{eq_dmi_berry} in terms of Green functions.
This can be achieved by first expressing the $\vn{k}$ and $\hat{\vn{n}}$
derivatives through the velocity and torque operators, respectively.
For this purpose we employ $\hbar\vn{v}=\partial H/\partial\vn{k}$
and $\hat{\vn{n}}\times\partial H/\partial \hat{\vn{n}}=\vn{\mathcal{T}}$
and use perturbation theory in order to rewrite the $\vn{k}$ and $\hat{\vn{n}}$
derivatives in terms of matrix elements of $\vn{v}$ 
and $\vn{\mathcal{T}}$~\cite{ibcsoit,phase_space_berry}.
This yields
\bege\label{eq_aknj_operator_form}
\vn{A}_{\vn{k}nj}=\hbar\sum_{m\neq n}\text{Im}
\left[
\frac{
\langle u_{\vn{k}n}  |\vn{\mathcal{T}}| u_{\vn{k}m}  \rangle
\langle u_{\vn{k}m}  |v_{j}| u_{\vn{k}n}  \rangle
}
{
\mathcal{E}_{\vn{k}m}-\mathcal{E}_{\vn{k}n}
}
\right]
\ee
and
\bege\label{eq_bknj_operator_form}
\vn{B}_{\vn{k}nj}
=-2\hbar\sum_{m\neq n}\text{Im}
\left[
\frac{
\langle u_{\vn{k}n}  |\vn{\mathcal{T}}| u_{\vn{k}m}  \rangle
\langle u_{\vn{k}m}  |v_{j}| u_{\vn{k}n}  \rangle
}
{
(\mathcal{E}_{\vn{k}m}-\mathcal{E}_{\vn{k}n})^2
}
\right],
\ee
where
\bege
\vn{\mathcal{T}}=
\frac{i}{2}
\left[
\bar{H}_{\vn{k}},
\vn{\sigma}
\right]=-\mu_{\rm B}\vn{\sigma}\times \hat{\vn{n}}\Bxc^{\rm
  xc}
\ee
is the torque 
operator~\cite{invsot}. 
Subsequently, we apply the identity
\bege
\begin{aligned}
&{\rm Im}
\int
\rmd\mathcal{E} g(\mathcal{E})
\frac{1}{\mathcal{E}-\mathcal{E}_{\vn{k}q}+i0^+}
\frac{1}{(\mathcal{E}-\mathcal{E}_{\vn{k}p}+i0^+)^2}
=\\
&=\pi
\left[
\frac{[g(\mathcal{E}_{\vn{k}p})-g(\mathcal{E}_{\vn{k}q})]}
{(\mathcal{E}_{\vn{k}p}-\mathcal{E}_{\vn{k}q})^2}
+
\frac{f(\mathcal{E}_{\vn{k}p})}
{\mathcal{E}_{\vn{k}q}-\mathcal{E}_{\vn{k}p}}
\right],
\end{aligned}
\ee
which can be proven with the residue theorem and with the
relation $g'(\mathcal{E})=f(\mathcal{E})$.
Thereby, we obtain the following expression
of the DMI coefficients in terms of Green functions~\cite{note_on_K_vs_F}: 
\begin{gather}\label{eq_dmi_green_functions_gder}
\begin{aligned}
&\vn{D}_{j}=\frac{1}{h}{\rm Re}
\int \frac{\rmd ^d k  }{(2\pi)^d}
\int \rmd\mathcal{E} 
g(\mathcal{E})\times\\
&\times{\rm Tr}
\left[
\vn{\mathcal{T}}
G_{\vn{k} }^{\rm R}(\mathcal{E})
v_{j}
\frac{dG_{\vn{k} }^{\rm R}(\mathcal{E})}{d\mathcal{E}}
-
\vn{\mathcal{T}}\frac{dG_{\vn{k} }^{\rm R}(\mathcal{E})}{d\mathcal{E}}v_{j}G_{\vn{k} }^{\rm R}(\mathcal{E})
\right],
\end{aligned}\raisetag{2\baselineskip}
\end{gather}
where $G_{\vn{k} }^{\rm
  R}(\mathcal{E})=\hbar\left[\mathcal{E}-H_{\vn{k}}+i0^{+}\right]^{-1}$
is the retarded Green function.

In order to identify the contributions  
to Eq.~\eqref{eq_dmi_green_functions_gder}
that are linear in SOI 
we expand $G_{\vn{k} }^{\rm R}(\mathcal{E})$ up to 
first order in $\mathcal{V}_{\vn{k}}$ as follows:
\bege\label{eq_green_first_order_soi}
G^{\rm R}_{\vn{k} } (\mathcal{E})\simeq
\bar{G}^{\rm R}_{\vn{k} } (\mathcal{E})+\frac{1}{\hbar}
\bar{G}^{\rm R}_{\vn{k} } (\mathcal{E})
\mathcal{V}_{\vn{k}}
\bar{G}^{\rm R}_{\vn{k} } (\mathcal{E}),
\ee
where $\bar{G}_{\vn{k} }^{\rm
  R}(\mathcal{E})=\hbar\left[\mathcal{E}-\bar{H}_{\vn{k}}+i0^{+}\right]^{-1}$
is the retarded Green function without SOI.
Additionally, we decompose the velocity operator into two parts:
\bege\label{eq_decompose_velocity}
v_{j}=\bar{v}_{j}+v^{\rm SOI}_{j},
\ee
where $\bar{v}_{j}=i[\bar{H}_{\vn{k}},r_j]/\hbar$ 
and $v^{\rm SOI}_{j}=i[\mathcal{V}_{\vn{k}},r_j]/\hbar$.
Inserting Eq.~\eqref{eq_green_first_order_soi} 
and Eq.~\eqref{eq_decompose_velocity} 
into Eq.~\eqref{eq_dmi_green_functions_gder}
and using the 
relation $\hbar \partial\bar{G}^{\rm R}_{\vn{k} }
(\mathcal{E})/\partial \mathcal{E}=-[\bar{G}^{\rm R}_{\vn{k} }(\mathcal{E})]^2$ we 
obtain the linear contribution of SOI
to $\vn{D}_{j}$:
\begin{gather}\label{eq_dmi_green_functions_soipertur}
\begin{aligned}
\vn{D}^{(1)}_{j}=&\frac{-1}{2\pi\hbar^3}
{\rm Re}
\int \frac{\rmd ^d k  }{(2\pi)^d}
\int \rmd\mathcal{E} 
g(\mathcal{E})
{\rm Tr}\Biggl[\\
&\gbar \vso
\gbar
\vn{\mathcal{T}}
\gbar
\bar{v}_{j}
\gbar\\
+& \gbar
\vn{\mathcal{T}}
\gbar \vso
\gbar
\bar{v}_{j}
\gbar\\
+&\gbar
\vn{\mathcal{T}}
\gbar
\bar{v}_{j}
\gbar \vso
\gbar\\
+&
\hbar
\gbar
\vn{\mathcal{T}}
\gbar
v^{\rm SOI}_{j}
\gbar\\
-&
\gbar \vso
\gbar
\bar{v}_{j}
\gbar
\vn{\mathcal{T}}
\gbar\\
-&
\gbar
\bar{v}_{j}
\gbar \vso
\gbar
\vn{\mathcal{T}}
\gbar\\
-&
\gbar
\bar{v}_{j}
\gbar
\vn{\mathcal{T}}
\gbar \vso
\gbar\\
-&
\hbar
\gbar
v^{\rm SOI}_{j}
\gbar
\vn{\mathcal{T}}
\gbar
\Biggr].
\end{aligned}\raisetag{2\baselineskip}
\end{gather}
Substituting the torque operator by the commutator of
inverse Green function and Pauli matrices, i.e.,
\bege\label{eq_torque_soi_comm}
\vn{\mathcal{T}}=-
\frac{i\hbar}{2}
\left[
[\gbar]^{-1},\vn{\sigma}
\right],
\ee
reduces the number of Green functions in each of the
products by one and
leaves us with
\begin{gather}\label{eq_dmi_green_functions_torquecommu}
\begin{aligned}
&\vn{D}^{(1)}_{j}=\frac{-1}{4\pi\hbar^2}{\rm Im}
\int \frac{\rmd ^d k  }{(2\pi)^d}
\int
\rmd\mathcal{E} g(\mathcal{E})
{\rm Tr}\Biggl[\\
&\gbar 
\mathcal{V}_{\vn{k}}
\vn{\sigma}
\gbar
\bar{v}_{j}
\gbar
-\gbar 
\mathcal{V}_{\vn{k}}
\gbar
\vn{\sigma}
\bar{v}_{j}
\gbar\\
+&
\vn{\sigma}
\gbar 
\mathcal{V}_{\vn{k}}
\gbar
\bar{v}_{j}
\gbar
-\gbar
\vn{\sigma}
\mathcal{V}_{\vn{k}}
\gbar
\bar{v}_{j}
\gbar\\
+&
\vn{\sigma}
\gbar
\bar{v}_{j}
\gbar 
\mathcal{V}_{\vn{k}}
\gbar
-\gbar
\vn{\sigma}
\bar{v}_{j}
\gbar 
\mathcal{V}_{\vn{k}}
\gbar\\
+&
\hbar
\vn{\sigma}
\gbar
v^{\rm SOI}_{j}
\gbar
-
\hbar
\gbar
\vn{\sigma}
v^{\rm SOI}_{j}
\gbar\\
-&
\gbar 
\mathcal{V}_{\vn{k}}
\gbar
\bar{v}_{j}
\vn{\sigma}
\gbar
+
\gbar 
\mathcal{V}_{\vn{k}}
\gbar
\bar{v}_{j}
\gbar
\vn{\sigma}
\\
-&
\gbar
\bar{v}_{j}
\gbar 
\mathcal{V}_{\vn{k}}
\vn{\sigma}
\gbar+
\gbar
\bar{v}_{j}
\gbar 
\mathcal{V}_{\vn{k}}
\gbar
\vn{\sigma}
\\
-&
\gbar
\bar{v}_{j}
\vn{\sigma}
\gbar 
\mathcal{V}_{\vn{k}}
\gbar
+
\gbar
\bar{v}_{j}
\gbar
\vn{\sigma}
\mathcal{V}_{\vn{k}}
\gbar\\
-&
\hbar
\gbar
v^{\rm SOI}_{j}
\vn{\sigma}
\gbar
+
\hbar
\gbar
v^{\rm SOI}_{j}
\gbar
\vn{\sigma}
\Biggr].
\end{aligned}\raisetag{2\baselineskip}
\end{gather}
We first pick out all the terms from 
Eq.~\eqref{eq_dmi_green_functions_torquecommu} that
can be expressed in terms of the 
anticommutators $\{\vn{\sigma},\bar{v}_j\}$
and $\{\vn{\sigma},v^{\rm SOI}_j\}$, because
we are searching for a relation
of the form of Eq.~\eqref{eq_dmi_spincurr} 
between DMI and the
ground-state spin current and
according to Eq.~\eqref{eq_spincurr} this
spin-current is given in terms of such anticommutators.
The sum of these contributions is given by
\bege\label{eq_aj}
\begin{aligned}
\vn{a}_{j}=&\frac{1}{4\pi\hbar^2}
{\rm Im}
\intkspa
\intene
g(\mathcal{E})
\Trace
\Biggl[\\
&\{\vn{\sigma},\bar{v}_j\}
[\gbar]^2
\mathcal{V}_{\vn{k}}
\gbar+
\\
+&\{
\vn{\sigma},\bar{v}_j
\}
\gbar
\mathcal{V}_{\vn{k}}
[\gbar]^2
+\hbar\{
\vn{\sigma},v^{\rm SOI}_j
\}[\gbar]^2
\Biggr]=\\
=&\frac{1}{4\pi\hbar}
{\rm Im}
\intkspa
\intene
f(\mathcal{E})
\Trace
\Biggl[\\
&\{\vn{\sigma},\bar{v}_j\}
\gbar
\mathcal{V}_{\vn{k}}
\gbar
+\hbar\{
\vn{\sigma},v^{\rm SOI}_j
\}\gbar
\Biggr],
\end{aligned}
\ee
where we used integration by parts and the relations $\hbar \partial\bar{G}^{\rm R}_{\vn{k} }
(\mathcal{E})/\partial \mathcal{E}=-[\bar{G}^{\rm R}_{\vn{k}
}(\mathcal{E})]^2$
and $g'(\mathcal{E})=f(\mathcal{E})$.
Carrying out the energy integrations with the help of the residue theorem
we obtain $\vn{a}_{j}=\vn{a}_{j}^{\rm (I)}+\vn{a}_{j}^{\rm (II)}$ with
\bege\label{eq_aj_one}
\begin{aligned}
\vn{a}_{j}^{\rm (I)}=&-\frac{\hbar}{4}
{\rm Re}
\intkspa
\sum_{n,m}
\frac{
f(\ebar_{\vn{k}n})
-f(\ebar_{\vn{k}m})
}
{
\ebar_{\vn{k}n}
-
\ebar_{\vn{k}m}
}\times\\
&\quad\quad\quad\times\langle
\bar{u}_{\vn{k}n}
|
\{\vn{\sigma},\bar{v}_j\}
|
\bar{u}_{\vn{k}m}
\rangle
\langle
\bar{u}_{\vn{k}m}
|
\mathcal{V}_{\vn{k}}
|
\bar{u}_{\vn{k}n}
\rangle
\end{aligned}
\ee
and
\bege\label{eq_aj_two}
\begin{aligned}
\vn{a}_{j}^{\rm (II)}=&-\frac{\hbar}{4}
{\rm Re}
\intkspa
\sum_{n}
f(\ebar_{\vn{k}n})
\langle\bar{u}_{\vn{k}n}
|
\{\vn{\sigma},v^{\rm SOI}_j\}
|
\bar{u}_{\vn{k}n}\rangle,
\end{aligned}
\ee
where $\ebar_{\vn{k}n}$ 
and $|\bar{u}_{\vn{k}n}\rangle$
are eigenvalues and eigenfunctions of the
Hamiltonian $\bar{H}_{\vn{k}}$ without SOI, 
i.e., $\bar{H}_{\vn{k}}|\bar{u}_{\vn{k}n}\rangle=\ebar_{\vn{k}n}|\bar{u}_{\vn{k}n}\rangle$. 
Expanding the spin current $\vn{Q}_{j}$ in Eq.~\eqref{eq_spincurr} analogously to
the expansion in Eq.~\eqref{eq_expand_dmi} we find
\bege
\vn{a}_{j}=-\vn{Q}^{(1)}_{j},
\ee
where $\vn{Q}^{(1)}_{j}$ is the SOI-linear term in 
\bege
\vn{Q}_{j}=\xi \vn{Q}^{(1)}_{j}+\xi^2 \vn{Q}^{(2)}_{j}+\xi^3 \vn{Q}^{(3)}_{j}+\dots.
\ee
Thus, the SOI-linear contribution to DMI 
generally contains the ground-state spin current. 
Eq.~\eqref{eq_aj_two} differs from Eq.~\eqref{eq_spincurr}
by the minus sign, by the replacement of the wave 
functions $|u_{\vn{k}n}\rangle$
by those without SOI, i.e., $|\bar{u}_{\vn{k}n}\rangle$, and by the
replacement of the velocity operator by its SOI correction $v_{j}^{\rm SOI}$.
Since $v_{j}^{\rm SOI}$ is first order in SOI, it is obvious that 
Eq.~\eqref{eq_aj_two} contributes to $-\vn{Q}^{(1)}_{j}$.
The second contribution, Eq.~\eqref{eq_aj_one}, arises when
first order perturbation theory is used to add SOI to the wave 
functions $|\bar{u}_{\vn{k}n}\rangle$ without SOI and when
the spin current $\hbar\{\vn{\sigma},v_{j}\}/4$ is 
evaluated for these perturbed wave functions.

Next, we pick out all terms from Eq.~\eqref{eq_dmi_green_functions_torquecommu}
that can be expressed in terms of the 
commutators $[\mathcal{V}_{\vn{k}},\vn{\sigma}]$. The sum of these
contributions is given by
\bege\label{eq_bj}
\begin{aligned}
\vn{b}_{j}=&\frac{-1}{4\pi\hbar^2}
{\rm Im}
\intkspa
\intene
g(\mathcal{E})
\Trace\Biggl[\\
&\quad\quad\quad\quad\gbar
[\mathcal{V}_{\vn{k}},\vn{\sigma}]
\gbar
\bar{v}_{j}
\gbar\\
&\quad\quad\quad-\gbar
\bar{v}_{j}
\gbar
[\mathcal{V}_{\vn{k}},\vn{\sigma}]
\gbar\Biggr]=\\
=&\frac{1}{h}
{\rm Re}
\intkspa
\intene
g(\mathcal{E})
\Trace\Biggl[\\
&\vn{\mathcal{L}}_{\vn{k}}
\gbar
\bar{v}_{j}
\frac{\partial\gbar}{\partial\mathcal{E}}
-
\vn{\mathcal{L}}_{\vn{k}}
\frac{\partial\gbar}{\partial\mathcal{E}}
\bar{v}_{j}
\gbar
\Biggr],
\end{aligned}
\ee
where we defined $\vn{\mathcal{L}}_{\vn{k}}=-i[\vso,\vn{\sigma}]/2$.
Using the residue theorem to perform the energy integrations yields
\bege\label{eq_bj_from_a_and_b}
\vn{b}_{j}=
\intkspa
\Biggl[
f(\ebar_{\vn{k}n})
\vn{\mathcal{A}}_{\vn{k}nj}
-
g(\ebar_{\vn{k}n})
\vn{\mathcal{B}}_{\vn{k}nj}
\Biggr],
\ee
where
\bege\label{eq_mathcal_a}
\vn{\mathcal{A}}_{\vn{k}nj}=
\hbar
\sum_{m\ne n}
{\rm Im}
\frac{
\langle
\bar{u}_{\vn{k}n}
|
\vn{\mathcal{L}}_{\vn{k}}
|
\bar{u}_{\vn{k}m}
\rangle
\langle
\bar{u}_{\vn{k}m}
|
\bar{v}_{j}
|
\bar{u}_{\vn{k}n}
\rangle
}
{
\bar{\mathcal{E}}_{\vn{k}m}-\bar{\mathcal{E}}_{\vn{k}n}
}
\ee
and
\bege\label{eq_mathcal_b}
\vn{\mathcal{B}}_{\vn{k}nj}=
-2\hbar
\sum_{m\ne n}
{\rm Im}
\frac{
\langle
\bar{u}_{\vn{k}n}
|
\vn{\mathcal{L}}_{\vn{k}}
|
\bar{u}_{\vn{k}m}
\rangle
\langle
\bar{u}_{\vn{k}m}
|
\bar{v}_{j}
|
\bar{u}_{\vn{k}n}
\rangle
}
{
(\bar{\mathcal{E}}_{\vn{k}n}-\bar{\mathcal{E}}_{\vn{k}m})^2
}.
\ee
Since $|\bar{u}_{\vn{k}n}\rangle$ is an eigenstate of the Hamiltonian
$\bar{H}_{\vn{k}}$ without SOI, we 
can use
\bege\label{eq_LK_time_reversal}
\begin{aligned}
&\langle
\bar{u}_{-\vn{k}n}
|
\vn{\mathcal{L}}_{-\vn{k}}
|
\bar{u}_{-\vn{k}m}
\rangle
=-
\left[
\langle
\bar{u}_{\vn{k}n}
|
\vn{\mathcal{L}}_{\vn{k}}
|
\bar{u}_{\vn{k}m}
\rangle
\right]^{*}
\\
&\langle
\bar{u}_{-\vn{k}m}
|
\bar{v}_{j}
|
\bar{u}_{-\vn{k}n}
\rangle=-
\left[
\langle
\bar{u}_{\vn{k}m}
|
\bar{v}_{j}
|
\bar{u}_{\vn{k}n}
\rangle
\right]^{*}
\end{aligned}
\ee
in order to show that 
$\vn{\mathcal{A}}_{-\vn{k}nj}=-\vn{\mathcal{A}}_{\vn{k}nj}$
and
$\vn{\mathcal{B}}_{-\vn{k}nj}=-\vn{\mathcal{B}}_{\vn{k}nj}$.
Therefore,
the $\vn{k}$ integration in Eq.~\eqref{eq_bj} evaluates 
to zero, i.e., $\vn{b}_{j}=0$.
Eq.~\eqref{eq_bj_from_a_and_b} strongly resembles Eq.~\eqref{eq_dmi_berry}.
The essential differences are that in $\vn{\mathcal{A}}_{\vn{k}nj}$ 
and $\vn{\mathcal{B}}_{\vn{k}nj}$ $\vn{\mathcal{L}}_{\vn{k}}$ takes the
place of $\vn{\mathcal{T}}$ in $\vn{A}_{\vn{k}nj}$ and $\vn{B}_{\vn{k}nj}$ and
the wave functions, eigenenergies and the velocity operator are replaced by those without SOI,
i.e., by $|\bar{u}_{\vn{k}n}\rangle$, $\bar{\mathcal{E}}_{\vn{k}n}$ 
and $\bar{v}_{j}$, respectively. Interestingly, the operator $\vn{\mathcal{L}}_{\vn{k}}$
is sometimes used instead of the torque operator $\vn{\mathcal{T}}$, because 
for stationary states in collinear magnets the expectation 
values are the same~\cite{nonlocal_torque_operators_random_alloys}.
Clearly, when we replace  $\vn{\mathcal{L}}_{\vn{k}}$ by $\vn{\mathcal{T}}$
both Eq.~\eqref{eq_mathcal_a} and Eq.~\eqref{eq_mathcal_b} yield zero, because
without SOI both DMI and spin-orbit torque are zero. Since $\vn{\mathcal{T}}$ may be
replaced by $\vn{\mathcal{L}}_{\vn{k}}$ in collinear ferromagnets, 
both Eq.~\eqref{eq_mathcal_a} and Eq.~\eqref{eq_mathcal_b} yield zero also 
with  $\vn{\mathcal{L}}_{\vn{k}}$. This is an alternative explanation why $\vn{b}_{j}=0$, 
which does not make use of Eq.~\eqref{eq_LK_time_reversal}.

Next, we discuss all those remaining terms 
in Eq.~\eqref{eq_dmi_green_functions_torquecommu} 
that contain $v^{\rm SOI}_{j}$. They are given by
\bege
\begin{aligned}
\vn{c}_{j}=
&\frac{-1}{2\pi\hbar}
{\rm Im}\!\!
\intkspa\!\!
\intene
g(\mathcal{E})
\Trace\Biggl[
\vn{\sigma}
\gbar
v^{\rm SOI}_{j}
\gbar
\Biggr]=\\
=&\frac{\hbar}{2}
\intkspa\sum_{n,m}
\frac{g(\ebar_{\vn{k}n})-g(\ebar_{\vn{k}m})}
{\ebar_{\vn{k}n}-\ebar_{\vn{k}m}}
\times\\
&\quad\quad\times\langle
\bar{u}_{\vn{k}n}
|
\vn{\sigma}
|
\bar{u}_{\vn{k}m}
\rangle
\langle
\bar{u}_{\vn{k}m}
|
v_{j}^{\rm SOI}
|
\bar{u}_{\vn{k}n}
\rangle.
\end{aligned}
\ee
Using $\vn{\mathcal{T}}=
i
\left[
\bar{H}_{\vn{k}},
\vn{\sigma}
\right]/2$ we can
rewrite the terms $\vn{c}_{j}$ 
and $\vn{a}_{j}^{\rm (II)}$ as follows:
\bege
\vn{c}_{j}
=-\int
\frac{\rmd ^d k}{(2\pi)^{d}}
\sum_{n}
g(\ebar_{\vn{k}n})
\tilde{\vn{B}}_{\vn{k}nj}
\ee
and
\bege
\vn{a}_{j}^{\rm (II)}
=\int
\frac{\rmd ^d k}{(2\pi)^{d}}
\sum_{n}
 f(\ebar_{\vn{k}n})\tilde{\vn{A}}_{\vn{k}nj},
\ee
where
\bege\label{eq_atildeknj}
\tilde{\vn{A}}_{\vn{k}nj}=\hbar\sum_{m\neq n}\text{Im}
\left[
\frac{
\langle 
\bar{u}_{\vn{k}n}  
|\vn{\mathcal{T}}| 
\bar{u}_{\vn{k}m}  \rangle
\langle 
\bar{u}_{\vn{k}m}  
|v^{\rm SOI}_{j}| 
\bar{u}_{\vn{k}n}  \rangle
}
{
\ebar_{\vn{k}m}-
\ebar_{\vn{k}n}
}
\right]
\ee
and
\bege\label{eq_btildeknj}
\tilde{\vn{B}}_{\vn{k}nj}
=-2\hbar\sum_{m\neq n}\text{Im}
\left[
\frac{
\langle 
\bar{u}_{\vn{k}n}  
|\vn{\mathcal{T}}| 
\bar{u}_{\vn{k}m}  
\rangle
\langle 
\bar{u}_{\vn{k}m}  
|v^{\rm SOI}_{j}| 
\bar{u}_{\vn{k}n}  
\rangle
}
{
(\ebar_{\vn{k}m}-\ebar_{\vn{k}n})^2
}
\right].
\ee
Comparing these expressions to those 
of $\vn{A}_{\vn{k}nj}$ and $\vn{B}_{\vn{k}nj}$ 
in Eq.~\eqref{eq_aknj_operator_form}
and in Eq.~\eqref{eq_bknj_operator_form} 
shows that 
$| u_{\vn{k}n} \rangle$, $\mathcal{E}_{\vn{k}n}$
and $v_{j}$
are 
replaced by $| \bar{u}_{\vn{k}n} \rangle$, $\ebar_{\vn{k}n}$
and $v^{\rm SOI}_{j}$, respectively.
While the contribution 
from $\tilde{\vn{A}}_{\vn{k}nj}$ enters
$\vn{Q}_{j}^{(1)}$, the contribution from
$\tilde{\vn{B}}_{\vn{k}nj}$ does not.

Finally, the last remaining terms in 
Eq.~\eqref{eq_dmi_green_functions_torquecommu} 
are given by 
\bege
\begin{aligned}
\vn{d}_{j}=&\frac{-1}{2\pi\hbar^2}
{\rm Im}
\intkspa
\intene
g(\mathcal{E})
\Trace\Biggl[\\
&\quad\quad\quad\quad
\vn{\sigma}
\gbar
\mathcal{V}_{\vn{k}}
\gbar
\bar{v}_{j}
\gbar+\\
&\quad\quad\quad\quad
+
\vn{\sigma}
\gbar
\bar{v}_{j}
\gbar
\mathcal{V}_{\vn{k}}
\gbar\Biggr].
\end{aligned}
\ee
Using the relation
\bege
\bar{v}_{j}=
\frac{i}{\hbar}
\left[
\bar{H}_{\vn{k}},r_{j}
\right]=
-i
\left[
[\gbar]^{-1},r_{j}
\right]
\ee
we can reduce the number of Green functions in each
of the products by one. This yields
\bege
\begin{aligned}
\vn{d}_{j}=&\frac{1}{2\pi\hbar}
{\rm Im}
\intkspa
\intene
g(\mathcal{E})
\Trace\Biggl[\\
&\quad\quad\quad\quad
\vn{\sigma}
\gbar
\frac{i}{\hbar}
[\mathcal{V}_{\vn{k}},
r_{j}
]
\gbar+\\
&\quad\quad\quad
+\frac{i}{\hbar}
[
\vn{\sigma}
,r_{j}
]
\gbar
\mathcal{V}_{\vn{k}}
\gbar
\Biggr].
\end{aligned}
\ee
Substituting $i [\mathcal{V}_{\vn{k}},r_{j}]/\hbar=v_{j}^{\rm SOI}$
and $[\vn{\sigma},r_{j}]=0$ we obtain
\bege
\vn{d}_{j}=-\vn{c}_{j}.
\ee

Finally, the SOI-linear contribution to DMI is given by
\bege
\label{eq_dmi1_equal_spicu1}
\vn{D}_{j}^{(1)}=\vn{a}_{j}+\vn{b}_{j}+\vn{c}_{j}+\vn{d}_{j}=\vn{a}_{j}=-\vn{Q}^{(1)}_{j}.
\ee
Thus, Eq.~\eqref{eq_dmi_spincurr} is satisfied in first order of $\vso$, which is the
main result of this section.
An interesting corollary of Eq.~\eqref{eq_dmi1_equal_spicu1} is that ground-state spin currents
in non-magnetic materials cannot arise at the first-order in $\vso$, because there is no
DMI in non-magnetic systems. 
In Section~\ref{sec_ajII} 
we will explicitly show that in non-magnetic systems $\vn{Q}^{(1)}_{j}=0$.
However, in the presence of SOI 
ground-state spin currents are possible 
in noncentrosymmetric crystals
even when they are non-magnetic, but these
are generated by higher order terms in the $\vso$-expansion. 
\section{Interpretation of the contributions $\vn{a}_{j}^{\rm(I)}$ and $\vn{a}_{j}^{\rm(II)}$}
At first order in SOI, both the DMI coefficient vector $\vn{D}_{j}^{(1)}$ 
and the ground-state spin current  $\vn{Q}_{j}^{(1)}$ can be decomposed into
two contributions, $\vn{a}_{j}^{\rm(I)}$ and $\vn{a}_{j}^{\rm(II)}$, according to
\bege
\vn{D}_{j}^{(1)}=-\vn{Q}_{j}^{(1)}=\vn{a}_{j}^{\rm(I)}+\vn{a}_{j}^{\rm(II)},
\ee
where $\vn{a}_{j}^{\rm(I)}$ and $\vn{a}_{j}^{\rm(II)}$ are given in Eq.~\eqref{eq_aj_one}
and in Eq.~\eqref{eq_aj_two}, respectively. In the following we discuss these two contributions
in detail.
\subsection{The contribution $\vn{a}_{j}^{\rm(I)}$}
\label{sec_gauge}
DMI can be interpreted as a Doppler shift due to the ground-state 
spin current $\vn{Q}_{j}$~\cite{dmi_doppler_shift}. In this interpretation
SOI is built in from the start and the resulting spin current
interacts with the noncollinear magnetic texture resulting in an
energy shift. In the following we discuss a complementary 
interpretation of the contribution $\vn{a}_{j}^{\rm(I)}$ to
DMI, which emphasizes the role of the spin-orbit field.
In contrast to Ref.~\cite{dmi_doppler_shift} we do not include
SOI from the start but instead we will add it later.
The Kohn-Sham Hamiltonian of a spin spiral without SOI is given by
\bege\label{eq_ks_hamil_spiral}
\bar{\mathcal{H}}=-\frac{\hbar^2}{2m}\Delta
+V(\vn{r})
+\mu_{\rm B}\Omega^{\rm xc}(\vn{r})\hat{\vn{n}}_{c}(\vn{r})\cdot\vht{\sigma}.
\ee 
We consider the special case of a flat cycloidal spin spiral 
with spin-spiral wavenumber $q$
propagating along the $x$ direction.
Its magnetization direction is 
\bege\label{eq_spin_spiral_cycloid}
\hat{\vn{n}}_{\rm c}(\vn{r})=\hat{\vn{n}}_{\rm c}(x)=
\begin{pmatrix}
\sin(qx)\\
0\\
\cos(qx)
\end{pmatrix}.
\ee

The
term $\mu_{\rm B}\Omega^{\rm xc}(\vn{r})\hat{\vn{n}}_{c}(\vn{r})\cdot\vht{\sigma}$ in
the Hamiltonian in Eq.~\eqref{eq_ks_hamil_spiral} can be
brought into the more convenient 
form $\mu_{\rm B}\Omega^{\rm xc}(\vn{r})\sigma_{z}$ by the
transformation
\bege\label{eq_gauge_trafo}
U^{\dagger}(\vn{r})
\mu_{\rm B}\Omega^{\rm xc}(\vn{r}) \hat{\vn{n}}_{c}(\vn{r})\cdot\vht{\sigma}
U(\vn{r})
=\mu_{\rm B}\Omega^{\rm xc}(\vn{r})\sigma_{z},
\ee
where
\bege\label{eq_gauge_trafo_matrix}
U(\vn{r})=
\left(
\begin{array}{cc}
\cos(\frac{qx}{2}) &-\sin(\frac{qx}{2})\\
\sin(\frac{qx}{2}) &\cos(\frac{qx}{2})
\end{array}
\right)
\ee
for the cycloidal spin-spiral in Eq.~\eqref{eq_spin_spiral_cycloid}.
Under the transformation $U(\vn{r})$ the Hamiltonian in
Eq.~\eqref{eq_ks_hamil_spiral} turns 
into~\cite{modified_llg_bazaliy_jones_zhang,bruno_the_2004}
\bege
\begin{aligned}
\tilde{H}=&
U^{\dagger}(\vn{r})
\bar{\mathcal{H}}
U(\vn{r})
=\\
=&\frac{1}{2m}(\vn{p}+e \vn{A}^{\rm eff})^2
+V(\vn{r})
+\mu_{\rm B}\Omega^{\rm xc}(\vn{r})\sigma_{z}
+\mathcal{O}(q^2),
\end{aligned}
\ee
where $\vn{p}=-i\hbar\vn{\nabla}$ and
$\mathcal{O}(q^2)$ summarizes
terms of order $q^2$ that we neglect in the following and 
\bege
\vn{A}^{\rm eff}=-\frac{i\hbar}{e}U^{\dagger}(\vn{r})
\frac{\partial U(\vn{r})}{\partial \vn{r}}
\ee
is an effective vector potential.
For the cycloidal spin-spiral in Eq.~\eqref{eq_spin_spiral_cycloid}
we obtain
\bege
\vn{A}^{\rm eff}=-\frac{\hbar q}{2e}\sigma_{y}\hat{\vn{e}}_{x},
\ee
where $\hat{\vn{e}}_{x}$ is a unit vector pointing in $x$ direction.
Up to first order in $q$ the transformed Hamiltonian is given by
\bege\label{eq_htilde_final}
\tilde{H}_{\vn{k}}\simeq
\bar{H}_{\vn{k}}-\frac{\hbar}{4}q
\{
\sigma_{y},\bar{v}_{x}
\}
,
\ee
where $\bar{H}_{\vn{k}}=e^{-i\vn{k}\cdot\vn{r}}\bar{H}e^{i\vn{k}\cdot\vn{r}}$ is the 
crystal-momentum representation of the
Hamiltonian of the corresponding magnetically collinear system:
\bege\label{eq_hamilton_collinear}
\bar{H}=-\frac{\hbar^2}{2m}\Delta
+V(\vn{r})+\mu_{\rm B}\Omega^{\rm xc}(\vn{r})\sigma_{z}.
\ee

The spectrum of Eq.~\eqref{eq_htilde_final} agrees to the spectrum of
Eq.~\eqref{eq_ks_hamil_spiral} (up to the first order in $q$) and the
eigenvectors of Eq.~\eqref{eq_ks_hamil_spiral} can be obtained from
the eigenvectors of Eq.~\eqref{eq_htilde_final} via the 
unitary transformation Eq.~\eqref{eq_gauge_trafo_matrix}.
However, solving the eigenvalue problem of Eq.~\eqref{eq_htilde_final} is 
considerably easier than solving the eigenvalue problem
of Eq.~\eqref{eq_ks_hamil_spiral}, in particular when the
wave vector $q$ of the spin-spiral in Eq.~\eqref{eq_spin_spiral_cycloid} 
is small. 

We denote the eigenstates of $\bar{H}_{\vn{k}}$ 
by $\left|\bar{u}_{\vn{k}n}\right\rangle$ and the corresponding
eigenenergies by $\bar{\mathcal{E}}_{\vn{k}n}$.
When an electron in band $n$ at $k$-point $\vn{k}$ propagates 
along the spin-spiral of
Eq.~\eqref{eq_spin_spiral_cycloid}
it exerts the torque $\tilde{\vn{\Torque}}^{\phantom{k}}_{\vn{k}n}$ on the magnetization.
From first order perturbation theory, where the second term
on the right-hand side of Eq.~\eqref{eq_htilde_final} acts as perturbation, we
obtain
\bege\label{eq_torque_ab_band}
\tilde{\vn{\Torque}}_{\vn{k}n}^{\phantom{k}}=
\frac{
\hbar q
}{2}
\sum_{m\ne n}{\rm Re}
\frac{
\Bigl\langle
\bar{u}_{\vn{k}n}
\Bigl|
\tilde{\vn{\mathcal{T}}}
\Bigr|
\bar{u}_{\vn{k}m}
\Bigr\rangle
\Bigl\langle
\bar{u}_{\vn{k}m}
\Bigl|
\{
\sigma_{y},\bar{v}_{x}
\}
\Bigr|
\bar{u}_{\vn{k}n}
\Bigr\rangle
}
{\ebar_{\vn{k}n}-\ebar_{\vn{k}m}},
\ee
where
$\tilde{\vn{\mathcal{T}}}(\vn{r})=-\mu_{\rm B}\vht{\sigma}\times
\hat{\vn{e}}_{z}\Omega^{\rm xc}(\vn{r})$ is 
the torque operator of the collinear system 
described by Eq.~\eqref{eq_hamilton_collinear}.
Using $
\tilde{\mathcal{T}}_{j}=i
\left[
\bar{H}_{\vn{k}},\sigma_{j}
\right]/2
$
we can simplify Eq.~\eqref{eq_torque_ab_band}
into
\bege
\tilde{\vn{\Torque}}^{\phantom{k}}_{\vn{k}n}=
-\frac{\hbar q}{2}
\hat{\vn{e}}_{x}^{\phantom{k}}
s_{\vn{k}n}^{\phantom{k}}
\bar{v}_{\vn{k}n x}^{\phantom{k}},
\ee
where 
$\bar{v}_{\vn{k}n x}^{\phantom{k}}=\langle \bar{u}_{\vn{k}n}^{\phantom{k}}|\bar{v}_{x}^{\phantom{k}}|\bar{u}^{\phantom{k}}_{\vn{k}n}\rangle$ is
the
group velocity in $x$ direction and the spin index
$s_{\vn{k}n}^{\phantom{k}}=\langle \bar{u}^{\phantom{k}}_{\vn{k}n}|\sigma ^{\phantom{k}}_{z}|\bar{u}^{\phantom{k}}_{\vn{k}n}\rangle $ is
1 for minority electrons and $-1$ for majority electrons.
Rotating $\tilde{\vn{\Torque}}_{\vn{k}n}^{\phantom{k}}$ back into the original reference
frame we obtain
\bege\label{eq_tkn_orig_ref_frame}
\vn{\Torque}^{\phantom{k}}_{\vn{k}n}=-\frac{\hbar q}{2}
s_{\vn{k}n}^{\phantom{k}}
\bar{v}_{\vn{k}n x}^{\phantom{k}}
\begin{pmatrix}
\cos(qx)\\
0\\
-\sin(qx)\\
\end{pmatrix},
\ee 
i.e., $\vn{\Torque}^{\phantom{k}}_{\vn{k}n}$ lies in the $zx$ plane and stays always 
perpendicular to $\hat{\vn{n}}_{c}(\vn{r})$
(Eq.~\eqref{eq_spin_spiral_cycloid}) while 
it rotates in the same
sense as $\hat{\vn{n}}_{c}(\vn{r})$.

The sign of the torque $\vn{\Torque}^{\phantom{k}}_{\vn{k}n}$ depends on the sign of 
the spin index $s^{\phantom{k}}_{\vn{k}n}$ as well as
on the sign of the group velocity $\bar{v}^{\phantom{k}}_{\vn{k}n x}$.
Since $\bar{v}^{\phantom{k}}_{-\vn{k}n x}=-\bar{v}^{\phantom{k}}_{\vn{k}n x}$
the Brillouin zone integral of $\vn{\Torque}^{\phantom{k}}_{\vn{k}n}$ is zero:
\bege\label{eq_fermi_sea_zero}
\intkspa\sum_{n}
f_{\vn{k}n}^{\phantom{k}}
\vn{\Torque}^{\phantom{k}}_{\vn{k}n}=0.
\ee
However, when an electric field is applied along $x$ direction,
a net torque on the magnetization of the spin-spiral
arises from the $\vn{\Torque}^{\phantom{k}}_{\vn{k}n}$, the so-called
adiabatic spin-transfer torque~\cite{modified_llg_bazaliy_jones_zhang,
Tatara_microscopic_approach_current_driven_domain_wall_dynamics,cit_metals_Haney_jmmm}.
If we approximate the relaxation time by the constant $\tau$,
the adiabatic spin-transfer torque, which we denote by $\vn{\Torque}^{\rm adia}$, is obtained
from Eq.~\eqref{eq_fermi_sea_zero} by considering that in the presence of the electric
field the occupancies $f_{\vn{k}n}$ change 
by $\delta f ^{\phantom{k}}_{\vn{k}n}=-e\tau \bar{v} ^{\phantom{k}}_{\vn{k}n x}
\delta(\mathcal{E}^{\phantom{k}}_{F}-\ebar^{\phantom{k}}_{\vn{k}n})E^{\phantom{k}}_x$.
We obtain
\bege\label{eq_adiabatic_torque}
\begin{aligned}
\vn{\Torque}^{\rm adia}=&-e\tau
\intkspa
\sum_{n}
\vn{\Torque}^{\phantom{k}}_{\vn{k}n}
\bar{v}^{\phantom{k}}_{\vn{k}n x}
\delta(\mathcal{E}^{\phantom{k}}_{F}-\ebar^{\phantom{k}}_{\vn{k}n})
E^{\phantom{k}}_{x}=\\
&\!\!\!\!\!\!\!\!\!\!\!\!\!\!\!\!=\frac{\hbar e\tau q}{2}
\intkspa
\sum_{n}
\!\!s^{\phantom{k}}_{\vn{k}n}
\bar{v}^2_{\vn{k}n x}\!\!
\begin{pmatrix}
\cos(qx)\\
0\\
-\sin(qx)\\
\end{pmatrix}
\!\!\delta(\mathcal{E}^{\phantom{k}}_{F}-\ebar^{\phantom{k}}_{\vn{k}n})
E^{\phantom{k}}_{x}\\
&\!\!\!\!\!\!\!\!\!\!\!\!\!\!\!\!=\frac{\hbar q}{2e}
(\sigma_{\uparrow}-\sigma_{\downarrow})
\begin{pmatrix}
\cos(qx)\\
0\\
-\sin(qx)\\
\end{pmatrix}
E_{x}\\
&\!\!\!\!\!\!\!\!\!\!\!\!\!\!\!\!=\frac{\hbar}{2e}
PJ_{x}
\frac{\partial\hat{\vn{n}}(\vn{r})}{\partial x}
=\frac{
\partial \vn{Q}_{x}(\vn{r})
}
{\partial x}
\end{aligned}
\ee 
where
\bege
P=
\frac{
\sigma_{\uparrow}-\sigma_{\downarrow}
}
{
\sigma_{\uparrow}+\sigma_{\downarrow}
}
\ee
is the polarization of the electric current,
\bege
\vn{Q}_{x}(\vn{r})=\frac{\hbar}{2e}
PJ_{x}\hat{\vn{n}}(\vn{r})
\ee
is the spin current density,
$\sigma_{\uparrow}$ and $\sigma_{\downarrow}$  are the respective contributions 
of the minority and majority electrons to
the
electrical conductivity and $J_{x}$ is the electrical current density.
The observation that Eq.~\eqref{eq_adiabatic_torque}, 
i.e., $\vn{\Torque}^{\rm adia}=\partial \vn{Q}_{x}(\vn{r})/\partial x$,
is the well-known expression for the adiabatic 
spin-transfer torque~\cite{modified_llg_bazaliy_jones_zhang,cit_metals_Haney_jmmm,Tatara_microscopic_approach_current_driven_domain_wall_dynamics} validates
the approach of combining the gauge transformation Eq.~\eqref{eq_gauge_trafo_matrix} with
first-order perturbation theory in Eq.~\eqref{eq_torque_ab_band} to
obtain $\vn{\Torque}^{\phantom{k}}_{\vn{k}n}$.

\begin{figure}
\includegraphics[width=\linewidth]{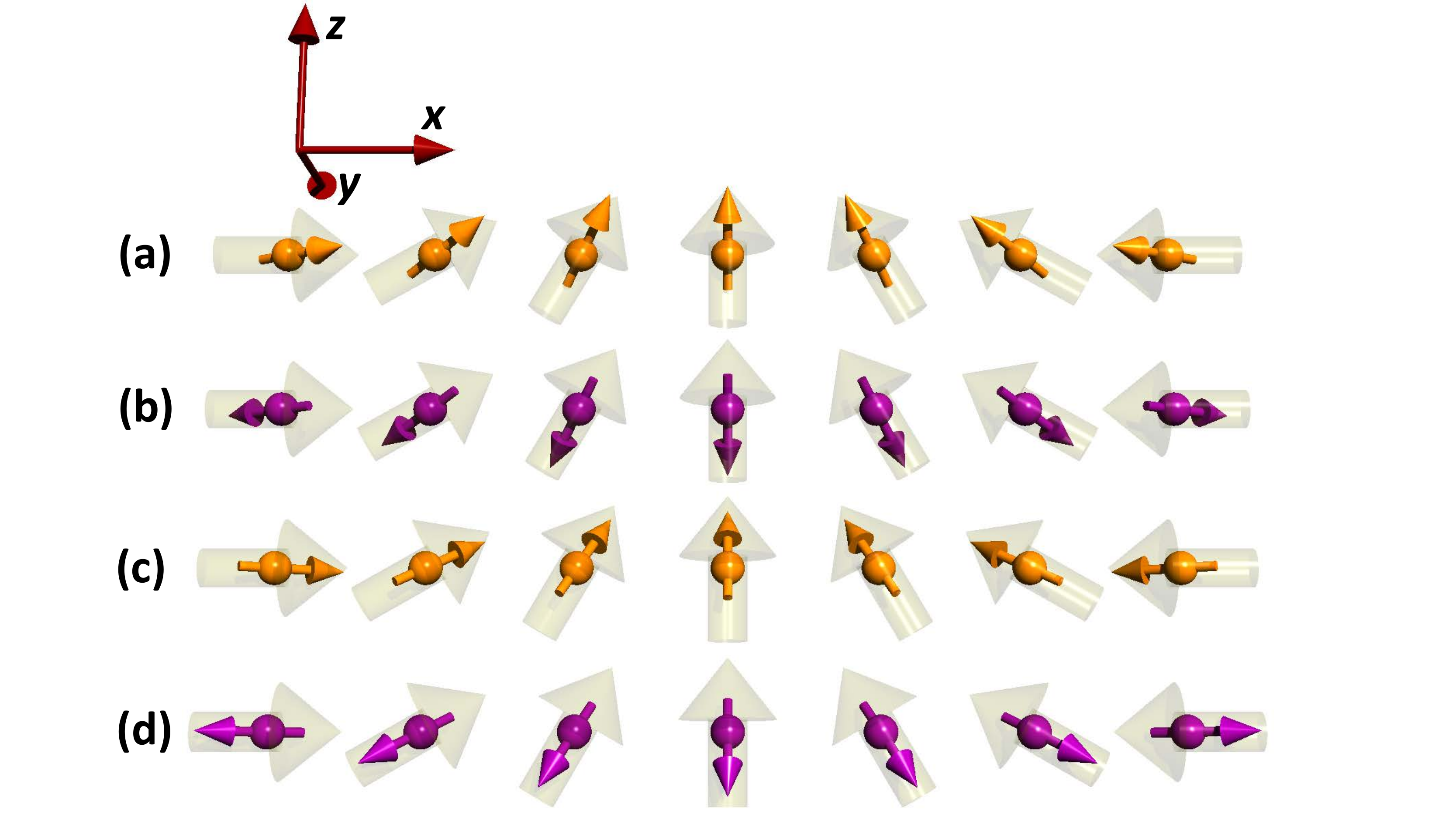}
\caption{\label{figure3}
Thick transparent arrows parallel to the $zx$ plane 
illustrate the magnetization direction
in the cycloidal spin spiral 
of Eq.~\eqref{eq_spin_spiral_cycloid} (for $q<0$).
Spheres with attached arrows
illustrate conduction electrons and their spins.
According to Eq.~\eqref{eq_conduc_elec_spin} the spin of a
conduction electron has a component in $y$ direction and is therefore
not perfectly aligned with the magnetization.   
(a) Minority electron with $v_{\vn{k}nx}<0$: $S_{\vn{k}ny}<0$.
(b) Majority electron with $v_{\vn{k}nx}<0$: $S_{\vn{k}ny}>0$.
(c) Minority electron with $v_{\vn{k}nx}>0$: $S_{\vn{k}ny}>0$.
(d) Majority electron with $v_{\vn{k}nx}>0$: $S_{\vn{k}ny}<0$.
}
\end{figure}

The origin of the torque $\vn{\Torque}^{\phantom{k}}_{\vn{k}n}$ is the $y$
component $S^{\phantom{k}}_{\vn{k}ny}$ of the spin that 
electrons acquire as they move along
the cycloidal spin-spiral of Eq.~\eqref{eq_spin_spiral_cycloid}:
\bege\label{eq_spin_in_spiral}
\begin{aligned}
S^{\phantom{k}}_{\vn{k}ny}&=-\frac{\hbar^2q}{4}
\sum_{m\ne n}{\rm Re}
\frac{
\Bigl\langle
\bar{u}_{\vn{k}n}
\Bigl|
\sigma_{y}
\Bigr|
\bar{u}_{\vn{k}m}
\Bigr\rangle
\Bigl\langle
\bar{u}_{\vn{k}m}
\Bigl|
\{
\sigma_{y},\bar{v}_{x}
\}
\Bigr|
\bar{u}_{\vn{k}n}
\Bigr\rangle
}
{\ebar_{\vn{k}n}-\ebar_{\vn{k}m}}\\
&=-\frac{\hbar^2q}{2}
\sum_{m\ne n}{\rm Im}
\frac{
\Bigl\langle
\bar{u}_{\vn{k}n}
\Bigl|
\tilde{\mathcal{T}}_{y}
\Bigr|
\bar{u}_{\vn{k}m}
\Bigr\rangle
\Bigl\langle
\bar{u}_{\vn{k}m}
\Bigl|
\{
\sigma_{y},\bar{v}_{x}
\}
\Bigr|
\bar{u}_{\vn{k}n}
\Bigr\rangle
}
{\left(\ebar_{\vn{k}n}-\ebar_{\vn{k}m}\right)^2}.\\
\end{aligned}
\ee
Assuming that the minority ($\uparrow$) and 
majority ($\downarrow$)
states differ only by
a rigid shift $\Delta\mathcal{E}$ of the bandenergies, i.e.,
\bege\label{eq_rigid_shift_model}
\begin{aligned}
\ebar_{\vn{k}n\uparrow}&=\ebar_{\vn{k}n}+\Delta\mathcal{E},\quad
|\bar{u}_{\vn{k}n\uparrow}\rangle=|\bar{u}_{\vn{k}n}\rangle
\begin{pmatrix}
1\\
0\\
\end{pmatrix},\\
\ebar_{\vn{k}n\downarrow}&=\ebar_{\vn{k}n},\quad
|\bar{u}_{\vn{k}n\downarrow}\rangle=|\bar{u}_{\vn{k}n}\rangle
\begin{pmatrix}
0\\
1\\
\end{pmatrix},\\
\end{aligned}
\ee
we can approximate Eq.~\eqref{eq_spin_in_spiral} by
\bege\label{eq_conduc_elec_spin}
S^{\phantom{k}}_{\vn{k}ny}\approx-s^{\phantom{k}}_{\vn{k}n}
\frac{\hbar^2q}{2\Delta\mathcal{E}}
\bar{v}^{\phantom{k}}_{\vn{k}nx}.
\ee
As in the case 
of $\vn{\Torque}^{\phantom{k}}_{\vn{k}n}$ (cf.~Eq.~\eqref{eq_tkn_orig_ref_frame}), 
the sign of
$S^{\phantom{k}}_{\vn{k}ny}$ depends not 
only on $s^{\phantom{k}}_{\vn{k}n}$ but also on the
sign of $\bar{v}^{\phantom{k}}_{\vn{k}nx}$. 
Therefore, as illustrated in Fig.~\ref{figure3}, electron spins with 
the same $s^{\phantom{k}}_{\vn{k}n}$ are tilted
out of the $zx$ plane in opposite directions if 
their $\bar{v}^{\phantom{k}}_{\vn{k}nx}$ differs
in sign.

We add now the effect of SOI, which is not taken into account
in Eq.~\eqref{eq_ks_hamil_spiral}.
In non-magnetic crystals with broken inversion symmetry 
the degeneracy between spin-up and spin-down bands 
is lifted by SOI, which can be
described by an
effective $\vn{k}$-dependent magnetic
field $\vn{\Omega}^{\rm SOI}_{\vn{k}n}$, which acts on the
electron spins
(see Refs.~\cite{rmp_fabian,book_winkler,
interplay_rashba_dresselhaus_review,rashba_review}
for reviews).
This so-called spin-orbit field is an odd function of $\vn{k}$
and may be expanded as~\cite{interplay_rashba_dresselhaus_review}
\bege\label{eq_soi_field}
\Omega^{\rm SOI}_{\vn{k}ni}=
\sum_{j}\chi^{(2)}_{nij}k_j^{\phantom{1}}+
\sum_{jlm}\chi^{(4)}_{nijlm}k_j^{\phantom{1}}k_l^{\phantom{1}}k_m^{\phantom{1}}+\dots
,
\ee
where $\chi_{nij}^{(2)}$ is the $ij$ element of an axial tensor of second rank, which
depends on the band index $n$.
$\chi_{nijlm}^{(4)}$ is the $ijlm$ element of an axial tensor of
fourth rank.

For the (001) and (111) surfaces of cubic transition metals such as Pt and W 
symmetry requires that axial second rank tensors be of the form 
\bege\label{eq_sym_cubic_surface}
\vht{\chi}_{n}^{(2)}=
\begin{pmatrix}
0 & \alpha_n & 0\\
-\alpha_n & 0 & 0 \\
0 & 0  & 0  \\
\end{pmatrix}
\ee 
if the coordinate frame is chosen such that the surface
normal is along $z$ direction.
The resulting Zeeman interaction between the spin-orbit field and the
electron spin is given by
\bege\label{eq_soi_approx_sof}
\mu^{\phantom{B}}_{\rm B}
\vn{\sigma}\cdot\vn{\Omega}^{\rm SOI}_{\vn{k}n}
=\mu^{\phantom{B}}_{\rm B}
\alpha^{\phantom{B}}_{n}
(\vn{\sigma}\times\vn{k})\cdot\hat{\vn{e}}^{\phantom{B}}_{z},
\ee
which has the form of the Rashba interaction with a band-dependent
Rashba parameter $\alpha^{\phantom{B}}_{n}$.

We now consider magnetic bilayers,
where a magnetic layer is deposited on the (001) or (111) surfaces of
cubic transition metals, such as Mn/W(001) and Co/Pt(111).
When the magnetization is described by the cycloidal spin spiral of
Eq.~\eqref{eq_spin_spiral_cycloid}, electrons travelling
in $x$ direction exhibit non-zero $y$ components of 
both spin (Eq.~\eqref{eq_conduc_elec_spin}) 
and spin-orbit field:
\bege\label{eq_s_and_l}
\begin{aligned}
S^{\phantom{k}}_{\vn{k}ny}&\approx
-s^{\phantom{k}}_{\vn{k}n}\frac{\hbar^2q}{2\Delta\mathcal{E}}
\bar{v}^{\phantom{k}}_{\vn{k}nx},\\
\vn{\Omega}_{\vn{k}n}^{\rm SOI}\cdot \hat{\vn{e}}^{\phantom{y}}_{y}&=
-\alpha_n^{\phantom{n}} k^{\phantom{x}}_{x}.
\end{aligned}
\ee
A linear-in-$q$ energy shift results from the
Zeeman interaction between $S^{\phantom{k}}_{\vn{k}ny}$
and $\vn{\Omega}_{\vn{k}n}^{\rm SOI}$: 
\bege\label{eq_dmi_approx1}
\Delta\mathcal{E}_{\rm DMI}^{\rm(I)}\approx
\frac{\mu_{\rm B}^{\phantom{D}}\hbar q}
{\Delta\mathcal{E}}\intkspa
\sum_{n}f ^{\phantom{D}}_{\vn{k}n} 
\alpha_n^{\phantom{n}} k^{\phantom{x}}_{x}
s^{\phantom{k}}_{\vn{k}n}
\bar{v}^{\phantom{k}}_{\vn{k}nx}.
\ee
We emphasize that the Brillouin zone integral 
of Eq.~\eqref{eq_conduc_elec_spin} is zero
because electrons with opposite $\vn{k}$ vectors have opposite
velocities $\bar{v}^{\phantom{k}}_{\vn{k}nx}$ and 
their $S^{\phantom{k}}_{\vn{k}ny}$ cancel. 
However, 
$\Delta\mathcal{E}_{\rm DMI}^{\rm (I)}$ is nonzero
since both $S^{\phantom{k}}_{\vn{k}ny}$ 
and $\vn{\Omega}_{\vn{k}n}^{\rm SOI}$
are odd functions of $\vn{k}$.
Eq.~\eqref{eq_s_and_l} and
Eq.~\eqref{eq_dmi_approx1}
are a central result of this section: They show how DMI is related to
the spin-orbit field and to the adiabatic spin-transfer torque.

While Eq.~\eqref{eq_dmi_approx1} provides a useful and intuitive
picture of the origin of DMI, it provides only a rather crude estimate
because we approximated SOI in the magnetic bilayer by 
Eq.~\eqref{eq_soi_approx_sof}
and we used 
Eq.~\eqref{eq_rigid_shift_model} to derive the approximation 
Eq.~\eqref{eq_conduc_elec_spin} of $S^{\phantom{k}}_{\vn{k}ny}$. 
In order to obtain a more accurate 
expression for $\Delta\mathcal{E}_{\rm DMI}^{\rm(I)}$ we
use the full spin-orbit interaction $H^{\rm SOI}$
instead of Eq.~\eqref{eq_dmi_approx1} and we do not
use the rigid shift model Eq.~\eqref{eq_rigid_shift_model}. This yields: 
\bege\label{eq_dmi_fop}
\begin{aligned}
&\Delta\mathcal{E}_{\rm DMI}^{\rm(I)}
\approx-\frac{\hbar q}{2}
\intkspa
\sum_{n}f_{\vn{k}n}
\sum_{m\ne n}{\rm Re}
\Biggl\{
\\
&\frac{
\Bigl\langle
\bar{u}_{\vn{k}n}
\Bigl|
\mathcal{V}_{\vn{k}}
\Bigr|
\bar{u}_{\vn{k}m}
\Bigr\rangle
\Bigl\langle
\bar{u}_{\vn{k}m}
\Bigl|
\{
\sigma_{y},\bar{v}_{x}
\}
\Bigr|
\bar{u}_{\vn{k}n}
\Bigr\rangle
}
{\ebar_{\vn{k}n}-\ebar_{\vn{k}m}}
\Biggr\}=\\
&=qa_{yx}^{\rm(I)}.
\end{aligned}
\ee
Here, $a_{yx}^{\rm(I)}$ is the $y$ component of $\vn{a}^{\rm(I)}_{x}$, 
i.e., $a_{yx}=\hat{\vn{e}}_{y}\cdot\vn{a}^{\rm(I)}_{x}$, 
where $\hat{\vn{e}}_{y}$ is the unit vector in $y$ direction. 
There are two ways to read this equation. The first way is to 
consider SOI, i.e., $\mathcal{V}_{\vn{k}}$, as perturbation. Then Eq.~\eqref{eq_dmi_fop}
describes the change of the 
observable $\hbar\{\sigma_{y},\bar{v}_{x}\}/4$ in response to the perturbation $\mathcal{V}_{\vn{k}}$, i.e., 
it describes part of the SOI-linear contribution to the ground-state spin current.
The second way to read Eq.~\eqref{eq_dmi_fop} has been described in detail in this section:
According to Eq.~\eqref{eq_htilde_final} we can consider $-\hbar q\{\sigma_y,\bar{v}_{x}\}/4$
as the perturbation arising from the noncollinear spin-spiral structure.
Then Eq.~\eqref{eq_dmi_fop} describes the response of the observable $\mathcal{V}_{\vn{k}}$ to the
noncollinear spin-spiral structure. The observable $\mathcal{V}_{\vn{k}}$ measures the Zeeman interaction
between the spin-orbit field and the noncollinearity-induced spin, i.e., an energy-shift due to DMI.
Eq.~\eqref{eq_dmi_fop} is a central result of this section:
The two ways of reading Eq.~\eqref{eq_dmi_fop} explain in a simple and intuitive
way why DMI and ground-state spin currents are related. 

In the discussion above we considered the special case of flat cycloidal spirals. 
For a general noncollinear magnetic texture an electron moving with 
velocity $\bar{\vn{v}}_{\vn{k}n}$ is
misaligned with the local magnetization by 
\bege\label{eq_spin_misalign}
\begin{aligned}
\vn{S}_{\vn{k}n}&\approx -s_{\vn{k}n}
\frac{\hbar^2}{2\Delta\mathcal{E}}
\hat{\vn{n}}\times(\bar{\vn{v}}_{\vn{k}n}\cdot\vn{\nabla})\hat{\vn{n}}=\\
&=
-\sum_{j}\bar{v}_{\vn{k}n j}
s_{\vn{k}n}
\frac{\hbar^2}{2\Delta\mathcal{E}}
\left[
\hat{\vn{n}}\times
\frac{
\partial \hat{\vn{n}}
}{
\partial r_{j}
}
\right]
,
\end{aligned}
\ee
which is obtained by generalizing Eq.~\eqref{eq_conduc_elec_spin}.
Assuming that the $\vn{k}$-linear term in the spin-orbit field Eq.~\eqref{eq_soi_field}
dominates we can write the energy shift due to the Zeeman interaction of $\vn{S}_{\vn{k}n}$ 
with $\vn{\Omega}^{\rm SOI}_{\vn{k}n}$ as
\bege
\begin{aligned}
&\Delta\mathcal{E}_{\rm DMI}^{\rm(I)}\approx
-\frac{\mu_{\rm B}\hbar}{\Delta\mathcal{E}}
\intkspa
\sum_{n}f_{\vn{k}n}\times\\
&\quad\times\sum_{j}
s_{\vn{k}n}
\bar{v}_{\vn{k}nj}
\left[
\vn{\chi}_n^{(2)}
\vn{k}
\right]
\cdot
\left[
\hat{\vn{n}}\times
\frac{
\partial \hat{\vn{n}}
}{
\partial r_{j}
}
\right].
\end{aligned}
\ee
Comparing this expression to Eq.~\eqref{eq_first_order_free_energy}
leads to the approximation
\bege
\vn{a}_{j}^{\rm(I)}\approx
-\frac{\mu_{\rm B}\hbar}{\Delta\mathcal{E}}
\sum_{n}
\vn{\chi}_n^{(2)}
\hat{\vn{e}}_{j}
\intkspa
f_{\vn{k}n}
s_{\vn{k}n}
\bar{v}_{\vn{k}nj}k_{j},
\ee
from which it follows that the tensor 
$a_{ij}^{\rm(I)}$
has the same symmetry properties as the
tensors
$\chi^{(2)}_{nij}$. 
This  result can also be concluded directly from
symmetry arguments, because
$D_{ij}=\hat{\vn{e}}_{i}\cdot \vn{D}_{j}$ is an axial tensor of second
rank exactly like $\chi^{(2)}_{ij}$. 
Thus,
$D_{ij}$ is of the form of Eq.~\eqref{eq_sym_cubic_surface}
for (001) and (111) surfaces of cubic transition metals.
Also the torkance tensor that describes the spin-orbit torque is an axial tensor
of second rank. Therefore, the symmetry of $D_{ij}$ can be determined from the
symmetry of the magnetization-even component of the torkance tensor,
which has been determined for all crystallographic 
point groups~\cite{sot_ebert,Ciccarelli_NiMnSb,sot_afms_symmetry_zelezny}.
\subsection{The contribution $\vn{a}_{j}^{\rm(II)}$}
\label{sec_ajII}
In Eq.~\eqref{eq_htilde_final} we describe the perturbation to the
Hamiltonian due to the noncollinearity by the anticommutator of the
Pauli matrix with the velocity operator $\bar{\vn{v}}$, where the
effect of SOI is missing in $\bar{\vn{v}}$ .  
In the presence of SOI the term $-\hbar q\{\sigma_y,v_x^{\rm SOI}\}/4$
therefore needs to be added to the perturbation. In the first order 
this leads to the energy correction described by 
$\vn{a}_{j}^{\rm(II)}$ in
Eq.~\eqref{eq_aj_two}. 

The SOI-correction of the velocity operator can be written as
\bege\label{eq_vsoi}
\vn{v}^{\rm SOI}=\frac{i}{\hbar}
[\mathcal{V}_{\vn{k}},\vn{r}]
=\frac{\mu_{\rm B}}{2 m c^2 e}
\vn{\sigma}\times\vn{\nabla}V,
\ee
where we used that SOI is described by the last term
in Eq.~\eqref{eq_ks_hamil}. Eq.~\eqref{eq_vsoi}
allows us to rewrite the
anticommutators $\{\sigma_{i},v^{\rm SOI}_{j}\}$
occurring in Eq.~\eqref{eq_aj_two} as
\bege
\{\sigma_{i},v^{\rm SOI}_{j}\}=
-\frac{\mu_{\rm B}}{m c^2 e}\epsilon_{ijk}\frac{\partial V(\vn{r})}{\partial r_{k}},
\ee
where we used $\{\sigma_{i},\sigma_{j}\}=2\delta_{ij}$.
Finally, we obtain
\bege\label{eq_aj2_pot_grad}
\begin{aligned}
a^{\rm (II)}_{ij}&=
\hat{\vn{e}}_{i}\cdot\vn{a}^{\rm (II)}_{j}=\\
&=
\frac{\hbar\mu_{\rm B}}{4 m c^2 e}\epsilon_{ijk}
\intkspa
\sum_{n}
f(\ebar_{\vn{k}n})
\left\langle
\bar{u}_{\vn{k}n}
\right|
\frac{\partial V(\vn{r})}{\partial r_{k}}
\left|
\bar{u}_{\vn{k}n}
\right\rangle,
\end{aligned}
\ee
where $\hat{\vn{e}}_{i}$ is the unit vector in the $i$-th cartesian direction,
$\epsilon_{ijk}$ is the Levi-Civita symbol and $a_{ij}^{\rm(II)}$ is the $i$-th
cartesian component of $\vn{a}_{j}^{\rm(II)}$.
Thus, $a^{\rm (II)}_{ij}$ is directly proportional to the expectation value of
the gradient of the scalar effective potential. In centrosymmetric systems this
expectation value is zero.

In noncentrosymmetric systems Eq.~\eqref{eq_aj2_pot_grad} is generally
non-zero, 
even in non-magnetic systems.
However, in non-magnetic systems $\vn{Q}_{j}^{(1)}$ has to vanish, because
according to Eq.~\eqref{eq_dmi1_equal_spicu1} 
it is related to DMI, which is zero in non-magnetic systems.
We
therefore prove now
that $\vn{a}_{j}^{\rm(I)}$ and $\vn{a}_{j}^{\rm(II)}$ cancel out in non-magnetic systems.
In non-magnetic systems Eq.~\eqref{eq_aj_one} can be rewritten as
\bege\label{eq_aj_one_nonmag}
\begin{aligned}
\vn{a}_{yx}^{\rm (I)}=&\hat{\vn{e}}_{y}\cdot\vn{a}_{x}^{\rm(I)}=-\frac{\hbar}{4}
{\rm Re}
\intkspa
\sum_{n,m,s}
\frac{
f(\ebar_{\vn{k}ns})
-f(\ebar_{\vn{k}ms})
}
{
\ebar_{\vn{k}ns}
-
\ebar_{\vn{k}ms}
}\times\\
&\times\langle
\bar{u}_{\vn{k}ns}
|
\{\sigma_{y},\bar{v}_x\}
|
\bar{u}_{\vn{k}m-s}
\rangle
\langle
\bar{u}_{\vn{k}m-s}
|
\mathcal{V}_{\vn{k}}
|
\bar{u}_{\vn{k}ns}
\rangle\\
&=-\frac{\hbar}{2}
{\rm Im}
\intkspa
\sum_{n,m,s}s
\frac{
f(\ebar_{\vn{k}ns})
-f(\ebar_{\vn{k}ms})
}
{
\ebar_{\vn{k}ns}
-
\ebar_{\vn{k}ms}
}\times\\
&\times\langle
\bar{u}_{\vn{k}ns}
|
\bar{v}_x
|
\bar{u}_{\vn{k}ms}
\rangle
\langle
\bar{u}_{\vn{k}m-s}
|
\mathcal{V}_{\vn{k}}
|
\bar{u}_{\vn{k}ns}
\rangle\\
&=-\frac{\mu_{\rm B}}{2 e\hbar c^2}
{\rm Re}
\intkspa
\sum_{n,m,s}
f(\ebar_{\vn{k}ns})
\times\\
&\times
\frac{
\partial
\langle
\bar{u}_{\vn{k}ns}
|}
{
\partial k_x
}
|
\bar{u}_{\vn{k}ms}
\rangle
\langle
\bar{u}_{\vn{k}ms}
|
\frac{\partial V(\vn{r})}{\partial r_z}
\frac{\partial \bar{H}_{\vn k}}{\partial k_x}
|
\bar{u}_{\vn{k}ns}
\rangle\\
&=\frac{\hbar\mu_{\rm B}}{4 e m c^2}
{\rm Re}
\intkspa
\sum_{n,s}
f(\ebar_{\vn{k}ns})
\times\\
&\times
\langle
\bar{u}_{\vn{k}ns}
|
\frac{\partial V(\vn{r})}{\partial r_z}
|
\bar{u}_{\vn{k}ns}
\rangle
.
\end{aligned}
\ee
To simplify the equations we
treat only the $y$ component
of $\vn{a}^{\rm(I)}_{x}$ in Eq.~\eqref{eq_aj_one_nonmag}.
The other components can be worked out analogously.
$s=\pm 1$ labels the spin, i.e., $s=+1$ denotes spin-up and $s=-1$
denotes
spin-down and we used
$\langle s |\sigma_{y}|-s\rangle=-is$.
Since the system is supposed to be non-magnetic, all energy levels are
at least doubly degenerate: $\ebar_{\vn{k}n+1}=\ebar_{\vn{k}n-1}$.
In the last step we have used integration by parts and we have
substituted the second $k_{x}$ derivative of $\bar{H}_{\vn{k}}$
by $\hbar^2/m$. Comparison of Eq.~\eqref{eq_aj2_pot_grad}
and Eq.~\eqref{eq_aj_one_nonmag}
shows that $a_{xy}^{\rm (I)}+a_{xy}^{\rm (II)}=0$ in non-magnetic
systems (Note that
in Eq.~\eqref{eq_aj_one_nonmag} the band index $n$ runs over doubly
degenerate states and there is an additional spin index $s$. In
Eq.~\eqref{eq_aj2_pot_grad} there is only one band index, which runs over
both spin-up and spin-down states).

\section{Interpretation of the ground-state energy current
  associated with DMI}
\label{sec_interpretation_ground_state_energy_current}
When the magnetization is time-dependent, e.g., when skyrmions or
domain walls are moving or when the magnetization is precessing
at the ferromagnetic resonance, a ground-state energy current
arises from DMI~\cite{itsot}. It is given by 
\bege
\mathscr{J}^{\rm DMI}_{j}=-\vn{D}_{j}
\cdot
\left(
\hat{\vn{n}}\times\frac{\partial \hat{\vn{n}}}{\partial t}
\right).
\ee
In Eq.~\eqref{eq_dmi_approx1} we model DMI by the
Zeeman interaction between the spin-orbit field and the
misalignment of the spins of conduction electrons with the
noncollinear magnetic texture. 
Based on this model we develop
an interpretation of the ground-state energy 
current $\mathscr{J}^{\rm DMI}_{j}$ in the following.

According to Eq.~\eqref{eq_spin_misalign} the
misalignment of the spin of the electron of band $n$ at
$k$ point $\vn{k}$ can be written as
\bege\label{eq_spin_misalign_general}
\vn{S}_{\vn{k}n}(\vn{r})=
\sum_{j}\mathscr{S}^{(\vn{k}n)}_{j}(\hat{\vn{n}}(\vn{r}))
\left(
\hat{\vn{n}}(\vn{r})
\times
\frac{\partial \hat{\vn{n}}(\vn{r})}
{\partial r_j}
\right),
\ee
where we introduced the misalignment 
coefficients $\mathscr{S}^{(\vn{k}n)}_{j}$, which are
given approximately by
\bege\label{eq_misalignment_coefficients}
\mathscr{S}^{(\vn{k}n)}_{j}
\approx
-s_{\vn{k}n}\frac{\hbar^2}{2\Delta\mathcal{E}}
\bar{v}_{\vn{k}nj}.
\ee
We now consider a magnetic texture $\hat{\vn{n}}(\vn{r},t)$
which moves with
velocity $\vn{w}$ such that 
\bege
\hat{\vn{n}}(\vn{r},t)=\vn{n}_{0}(\vn{r}-\vn{w}t)
\ee
where $\vn{n}_{0}(\vn{r})$ describes for example a domain-wall or
a skyrmion at rest.
For such a magnetic texture in motion
the spin misalignment is
time-dependent and satisfies the continuity equation
\bege\label{eq_continuity_gssc}
\frac{\partial \vn{S}_{\vn{k}n}}{\partial t}+
\sum_{j}
\frac{\partial \vn{Q}^{(\vn{k}n)}_{j}}{\partial r_{j}}=0,
\ee
where
\bege\label{eq_gssc}
\vn{Q}^{(\vn{k}n)}_{j}=-
\mathscr{S}^{(\vn{k}n)}_{j}(\hat{\vn{n}})
\left(
\hat{\vn{n}}
\times
\frac{\partial \hat{\vn{n}}}
{\partial t}
\right)
\ee
is the misaligned part of the
spin-current density 
driven by the motion of the magnetic structure
and associated with the electron in band $n$
at $k$-point $\vn{k}$.  

The Brillouin zone integral of $\vn{Q}^{(\vn{k}n)}_{j}$ is zero:
\bege\label{eq_intkspa_spin_misalign}
\sum_{n}
\intkspa f(\ebar_{\vn{k} n})
\vn{Q}^{(\vn{k}n)}_{j}=0,
\ee
because according to Eq.~\eqref{eq_misalignment_coefficients}
the misalignment coefficients are odd functions 
of $\vn{k}$, i.e., $\mathscr{S}^{(-\vn{k}n)}_{j}=-\mathscr{S}^{(\vn{k}n)}_{j}$,
and consequently $\vn{Q}^{(\vn{k}n)}_{j}$ is also an odd function of $\vn{k}$, 
i.e., $\vn{Q}^{(-\vn{k}n)}_{j}=-\vn{Q}^{(\vn{k}n)}_{j}$.
Consequently, Eq.~\eqref{eq_gssc} does not lead to a net spin current, but it
describes counter-propagating spin-currents, where the spin-current 
carried by electrons at $k$-point $\vn{k}$ has the opposite sign of the
spin-current carried by electrons at $-\vn{k}$.

From Eq.~\eqref{eq_spin_misalign_general} 
until Eq.~\eqref{eq_intkspa_spin_misalign} SOI is not yet considered. Therefore,
the ground-state spin current associated with SOI is not present.
However, even in the absence of SOI, there are several additional spin
currents present
that are not included in our definition of $\vn{Q}^{(\vn{k}n)}_{j}$.
First, there is the spin current
\bege
\vn{Q}^{\rm xc}_{j}=
\sum_{n}
\intkspa f(\ebar_{\vn{k} n})
\vn{S}_{\vn{k}n}\bar{v}_{\vn{k}nj},
\ee
which mediates the exchange-stiffness torque. 
In contrast to $\vn{Q}^{(\vn{k}n)}_{j}$ the spin current
$\vn{Q}^{\rm xc}_{j}$ is zero in collinear systems, i.e, 
whenever $\partial\hat{\vn{n}}/\partial\vn{r}=0$. However,
the Brillouin-zone integration leads to a nonzero value 
of $\vn{Q}^{\rm xc}_{j}$ in noncollinear systems, while the
Brillouin-zone
integral of $\vn{Q}^{(\vn{k}n)}_{j}$ vanishes according 
to Eq.~\eqref{eq_intkspa_spin_misalign}. 
Second, the time-dependence of $\hat{\vn{n}}$ can lead to 
spin-pumping, in particular in magnetic bilayer systems.
Most discussions on spin-pumping focus on net 
spin currents (by 'net' we mean that in 
contrast to Eq.~\eqref{eq_intkspa_spin_misalign} the Brillouin zone
integral is not zero)
that flow 
in magnetic bilayer systems
from the magnet into the nonmagnet.
In contrast, the spin currents described by Eq.~\eqref{eq_gssc}   
are counter-propagating, 
i.e., $\vn{Q}^{(-\vn{k}n)}_{j}=-\vn{Q}^{(\vn{k}n)}_{j}$.
However, in analogy to the discussion of Eq.~\eqref{eq_s_and_l},
we will show now that such counter-propagating
spin currents are exactly what is needed to interpret the
ground-state energy current $\mathscr{J}^{\rm DMI}_{j}$.

In order to include SOI, we 
multiply Eq.~\eqref{eq_continuity_gssc} by
the spin-orbit field. Subsequently, we integrate over
the Brillouin zone and add the contributions of all occupied bands.
This yields the continuity equation for DMI energy
\bege\label{eq_continuity_dmi_energy}
\frac{\partial (\Delta \mathcal{E}^{\rm (I)}_{\rm DMI})}{\partial t}+
\sum_{j}\frac{
\partial 
\mathscr{J}_{j}^{\rm DMI,I}
}{\partial r_j}=0,
\ee
where 
\bege
\mathscr{J}_{j}^{\rm DMI,I}=
\frac{2\mu_{\rm B}}{\hbar}
\sum_{n}\intkspa
f(\ebar_{\vn{k}n})
\vn{Q}^{(\vn{k}n)}_{j}\cdot
\vn{\Omega}_{\vn{k}n}^{\rm SOI}
\ee
is a ground-state energy current associated with DMI.
This approximation leads to the
picture that $\mathscr{J}_{j}^{\rm DMI,I}$ is associated with counter-propagating
spin currents, where the spins carry energy due to their Zeeman
interaction with the spin-orbit field.
Since both $\vn{Q}^{(\vn{k}n)}_{j}$ 
and $\vn{\Omega}_{\vn{k}n}^{\rm SOI}$
are odd functions of $\vn{k}$, $\mathscr{J}_{j}^{\rm DMI,I}$ is nonzero in 
systems with inversion asymmetry. 

The continuity
equation of DMI energy,
Eq.~(82), has been discussed in detail
in Ref.~\cite{itsot}. It describes that DMI energy associated with
domain walls or skyrmions in motion moves together with these objects.
In this section we have shown that Eq.~\eqref{eq_continuity_dmi_energy} results
from the continuity equation, Eq.~\eqref{eq_continuity_gssc}, of the spin misalignment.

\section{Rashba model}
\label{sec_rashba_model}
We consider the model Hamiltonian
\bege\label{eq_rashba_model}
H_{\vn{k}}=\frac{\hbar^2}{2m}k^2+
\alpha (\vn{k}\times\hat{\vn{e}}_{z})\cdot\vn{\sigma}+
\frac{\Delta V}{2}\sigma_z,
\ee
where the first term is the kinetic energy,
the second term describes the Rashba spin-orbit coupling
and the third term describes the exchange interaction.
The velocity operators are given by the expressions~\cite{universal_she}
\bege
\begin{aligned}
v_{x}&=\frac{\hbar}{m}k_{x}-\frac{\alpha}{\hbar} \sigma_{y}\\
v_{y}&=\frac{\hbar}{m}k_{y}+\frac{\alpha}{\hbar} \sigma_{x}
\end{aligned}
\ee
and the spin velocity operators are~\cite{suppression_she}
\bege
\begin{aligned}
\frac{\hbar}{4}
\{
v_{y},\sigma_{x}
\}
=\frac{\hbar^2}{2m}k_{y}\sigma_{x}+\frac{1}{2}\alpha\\
\frac{\hbar}{4}
\{
v_{x},\sigma_{y}
\}
=\frac{\hbar^2}{2m}k_{x}\sigma_{y}-\frac{1}{2}\alpha.
\end{aligned}
\ee

\begin{figure}
\includegraphics[width=\linewidth,trim=0cm 0cm 6cm 18cm,clip]{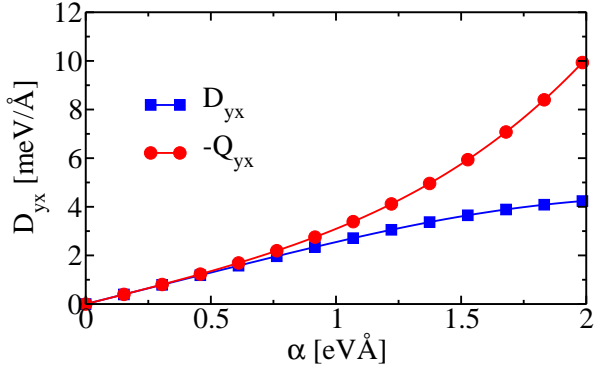}
\caption{\label{fig_rashba_model}
DMI coefficient $D_{yx}$ (squares) and
spin current $Q_{yx}$ (circles) vs.\ $\alpha$ in the
Rashba model Eq.~\eqref{eq_rashba_model} for the
parameters $\mu=0$ and $\Delta V=$1eV.
}
\end{figure}

First, we discuss the case $\Delta V=0$, where DMI vanishes.
For the ground-state spin currents, Eq.~\eqref{eq_spincurr}, the
following analytical expressions are readily derived
\bege\label{eq_rashba_spincurr}
\begin{aligned}
&\vn{Q}_{x}=-\frac{m^2\alpha^3}{6\pi\hbar^4}\hat{\vn{e}}_{y},\\
&\vn{Q}_{y}=\frac{m^2\alpha^3}{6\pi\hbar^4}\hat{\vn{e}}_{x},
\end{aligned}
\ee
when the temperature $T=0$ and when the chemical potential $\mu>0$.
In agreement with our discussion in section~\ref{sec_ajII} there is no
$\alpha$-linear term in Eq.~\eqref{eq_rashba_spincurr} because the system is
nonmagnetic due to $\Delta V=0$. On the other hand, the ground-state spin currents
are nonzero even in this nonmagnetic case. Since DMI vanishes for 
nonmagnetic systems Eq.~\eqref{eq_dmi_spincurr} is violated, while Eq.~\eqref{eq_dmi1_equal_spicu1} is
fulfilled.

Next, we discuss the magnetic case. Figure~\ref{fig_rashba_model} shows
the DMI-coefficient $D_{yx}$
and the ground-state spin current $Q_{yx}$
for
$\Delta V=$1eV at $\mu=0$ as a function of $\alpha$. In the SOI-linear regime, i.e., for small $\alpha$, 
we find $D_{yx}=-Q_{yx}$, in agreement with the analytical result in Eq.~\eqref{eq_dmi1_equal_spicu1}.
For large values of $\alpha$ 
nonlinearities become pronounced in
both $D_{yx}$ and $Q_{yx}$. Interestingly, $Q_{yx}$ exhibits stronger nonlinearity than $D_{yx}$ does.
Therefore, also in the magnetic Rashba 
model Eq.~\eqref{eq_dmi_spincurr} is violated, while Eq.~\eqref{eq_dmi1_equal_spicu1} is
fulfilled.

\section{Ab-initio calculations}
\label{sec_ab_initio}
In the following we discuss DMI and ground-state spin currents in
Mn/W(001) and Co/Pt(111) magnetic bilayers based on
\textit{ab-initio} density-functional theory calculations. 
The Mn/W(001) system consists of one monolayer of
Mn deposited on 9 layers of W(001) in our simulations. 
The Co/Pt(111) bilayer is composed of 3 layers of Co
deposited on 10 layers of Pt(111).
Computational details of the electronic structure calculations
are given in Ref.~\cite{ibcsoit}, where we investigated spin-orbit
torques in Mn/W(001) and Co/Pt(111).
We use Eq.~\eqref{eq_spincurr} to evaluate the full ground-state
spin-current density $Q_{yx}=\hat{\vn{e}}_{y}\cdot\vn{Q}_{x}$. Its
SOI-linear part $Q^{(1)}_{yx}=-a_{yx}^{\rm (I)}-a_{yx}^{\rm(II)}$
is calculated from Eq.~\eqref{eq_aj_one}
and from Eq.~\eqref{eq_aj_two}.
We employ Eq.~\eqref{eq_dmi_berry} to compute the DMI 
coefficient $D_{yx}=\hat{\vn{e}}_{y}\cdot\vn{D}_{x}$.
The temperature in the Fermi function $f(\mathcal{E})$ and in the
grand canonical potential density $g(\mathcal{E})$ is set 
to $T=300$K.

\begin{figure}
\includegraphics[width=\linewidth,trim=4.0cm 4.5cm 3.5cm 4.5cm,clip]{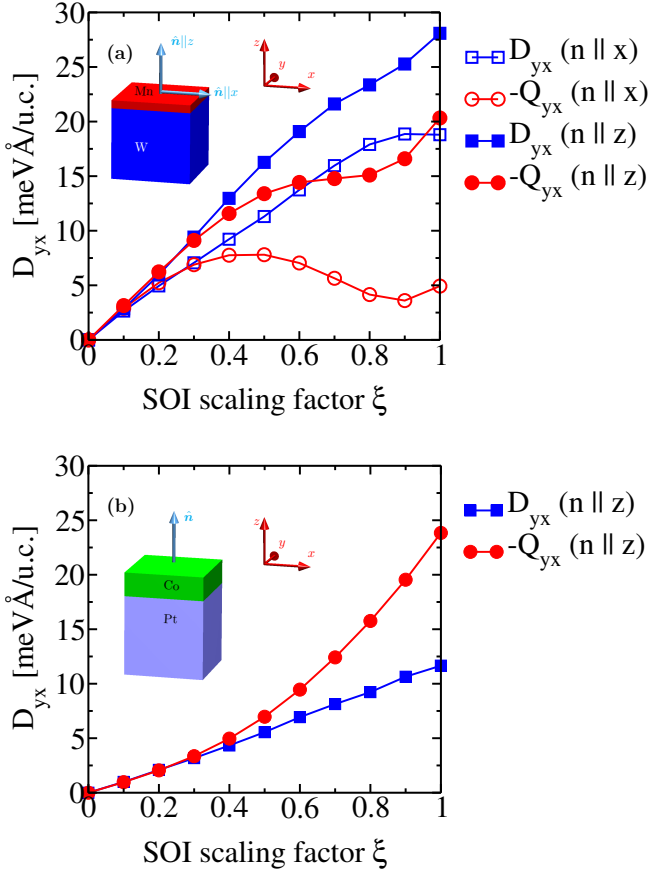}
\caption{\label{fig_compare_spicu_dmi_vs_xi}
DMI coefficient $D_{yx}$ (squares) and
spin current $Q_{yx}$ (circles) vs.\ SOI scaling factor $\xi$
in (a) Mn/W(001) and (b) Co/Pt(111). 
For the Mn/W(001) magnetic bilayer results are shown for magnetization along
$z$ ($\hat{\vn{n}}\Vert z$) and for magnetization 
along $x$ ($\hat{\vn{n}}\Vert x$). 
For the Co/Pt(111) magnetic bilayer results are shown for magnetization along
$z$.
The insets illustrate the geometry and the coordinate system.
}
\end{figure}

In Figure~\ref{fig_compare_spicu_dmi_vs_xi} we plot both the spin current
$Q_{yx}=\hat{\vn{e}}_{y}\cdot\vn{Q}_{x}$ and the DMI coefficient
$D_{yx}=\hat{\vn{e}}_{y}\cdot\vn{D}_{x}$ as a function of SOI scaling factor $\xi$
for the two systems Mn/W(001) and Co/Pt(111). The figure shows that in the linear
regime, i.e., for small $\xi$, the relation $D_{yx}=-Q_{yx}$ is satisfied very well, in agreement
with the analytical result in Eq.~\eqref{eq_dmi1_equal_spicu1}. For large values of $\xi$ nonlinear
contributions become important for both $Q_{yx}$ and $D_{yx}$ and the 
DMI coefficient is no longer
described very well by the ground-state spin current.

In Mn/W(001) we show $Q_{yx}$ and $D_{yx}$ for two different 
magnetization directions $\hat{\vn{n}}$, namely
$\hat{\vn{n}}\Vert z$ and $\hat{\vn{n}}\Vert x$. 
Both $Q_{yx}$ and $D_{yx}$ depend on $\hat{\vn{n}}$.
For small $\xi$ the $\hat{\vn{n}}$-dependence of $D_{yx}$ is well described by the 
$\hat{\vn{n}}$-dependence of $Q_{yx}$. However, at $\xi=1$ 
the $\hat{\vn{n}}$-dependence of $Q_{yx}$
is much stronger than 
the $\hat{\vn{n}}$-dependence of $D_{yx}$.

Interestingly, $Q_{yx}$ is much more 
nonlinear in $\xi$ than $D_{yx}$ in the considered $\xi$ range.
This leads to large deviations between $D_{yx}$ and $-Q_{yx}$ at $\xi=1$.
Since $D_{yx}$ is almost linear up to $\xi=1$, 
a good approximation is $D_{yx}\approx -Q^{(1)}_{yx}$.
This is a major difference to the B20 
compounds Mn$_{1-x}$Fe$_{x}$Ge 
and Fe$_{1-x}$Co$_{x}$Ge for which $D_{yx}\approx -Q_{yx}$ has been found to be a 
good approximation~\cite{dmi_doppler_shift}.
Due to the strong SOI from
the 5$d$ heavy metals in the
 magnetic bilayer systems considered in this work the SOI-nonlinear contributions in $Q_{yx}$
require to extract the SOI-linear part $Q^{(1)}_{yx}$ in order to approximate DMI by the
spin current as $D_{yx}\approx -Q^{(1)}_{yx}$.

\begin{figure}
\includegraphics[width=\linewidth,trim=4.0cm 4.5cm 3.5cm 4.5cm,clip]{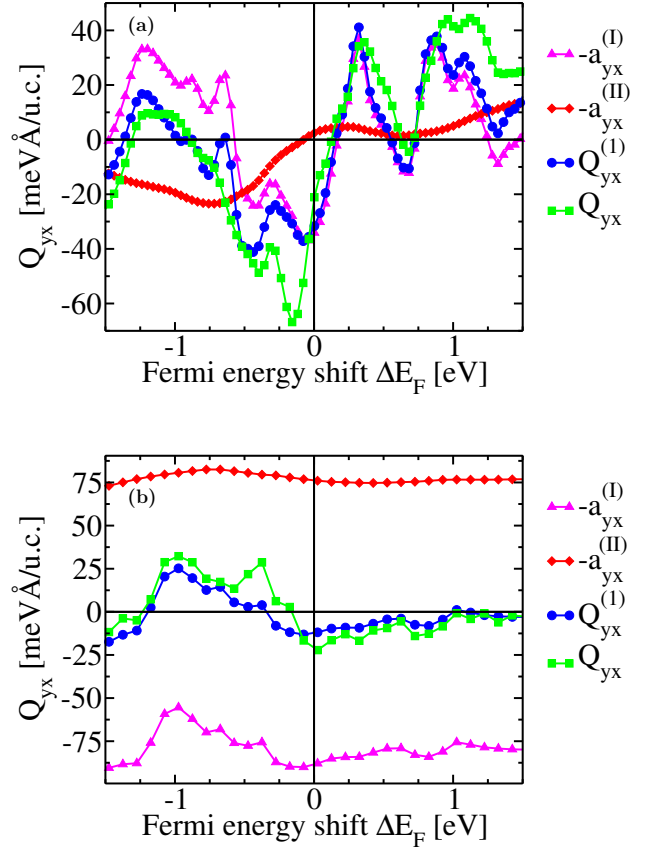}
\caption{\label{fig_decompose_spincurr}
Spin current $Q_{yx}$ (squares), SOI linear contribution $Q^{(1)}_{yx}$ to the spin current (circles),
and the contributions $a_{yx}^{\rm (I)}$ (triangles) and $a_{yx}^{\rm (II)}$ (diamonds) 
to $Q^{(1)}_{yx}=-a_{yx}^{\rm (I)}-a_{yx}^{\rm (II)}$.
Results are shown as a function of Fermi energy shift $\Delta E_{\rm F}$
in (a) Mn/W(001) and (b) Co/Pt(111) for magnetization in $z$ direction.
}
\end{figure}

The finding that $D_{yx}\approx -Q^{(1)}_{yx}$ is a good approximation 
despite the strong SOI from the heavy metals 
motivates us to investigate $Q^{(1)}_{yx}$ further by splitting it up
into its two contributions $-a_{yx}^{\rm (I)}$ and $-a_{yx}^{\rm(II)}$.
In Figure~\ref{fig_decompose_spincurr} we show $Q^{(1)}_{yx}$ and 
the two contributions $-a_{yx}^{\rm (I)}$ and $-a_{yx}^{\rm(II)}$.
In order to investigate chemical trends
we artificially shift the Fermi level by $\Delta E_{\rm F}$.
The Mn/W and Co/Pt bilayer systems correspond to $\Delta E_{\rm F}=0$. 
Negative values of $\Delta E_{\rm F}$ approximately describe the doped systems
Mn$_{1-x}$Cr$_{x}$/W$_{1-x}$Ta$_{x}$ and Co$_{1-x}$Fe$_{x}$/Pt$_{1-x}$Ir$_{x}$, while
positive values of $\Delta E_{\rm F}$ approximately describe the doped systems
Mn$_{1-x}$Fe$_{x}$/W$_{1-x}$Re$_{x}$ and Co$_{1-x}$Ni$_{x}$/Pt$_{1-x}$Au$_{x}$.
Figure~\ref{fig_decompose_spincurr} clearly shows that 
$a_{yx}^{\rm (I)}$ and $a_{yx}^{\rm(II)}$ are generally of similar 
magnitude. In Mn/W(001) at $\Delta E_{\rm F}=0$ there is a crossing of $\vn{a}_{yx}^{\rm(II)}$ 
through zero and therefore $Q^{(1)}_{yx}\approx -a_{yx}^{\rm(I)}$. However,
for $\Delta E_{\rm F}\ne 0$
both $a_{yx}^{\rm(I)}$ and $a_{yx}^{\rm(II)}$ are important.
In Co/Pt(111) both  $a_{yx}^{\rm(I)}$ and $a_{yx}^{\rm(II)}$ are much larger in magnitude
than $Q^{(1)}_{yx}$ but opposite in sign.
We also show $Q_{yx}$ in
Figure~\ref{fig_decompose_spincurr}. $Q_{yx}$ and $Q^{(1)}_{yx}$
behave similarly as a function of $\Delta E_{\rm F}$. However, due to
the strong SOI from the 5$d$ transition metal, $Q_{yx}$ and
$Q^{(1)}_{yx}$
often deviate substantially.

\begin{figure}
\includegraphics[width=\linewidth,trim=0cm 0cm 6cm 18cm,clip]{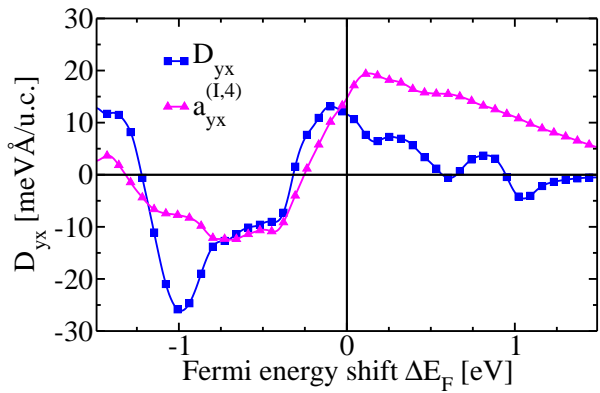}
\caption{\label{fig_contribution_aI4_vs_dmi}
DMI coefficient $D_{yx}$ (squares)
and the part $a_{yx}^{\rm (I,4)}$ of $a_{yx}^{\rm (I)}$ where 
SOI is switched-off in all Pt layers that are far away from the Co/Pt interface (triangles).
Results are shown as a function of Fermi energy shift $\Delta E_{\rm F}$
in Co/Pt(111) for magnetization in $z$ direction.
}
\end{figure}

Numerical tests in Fe and Ni have shown that the effect of the SOI-correction $v^{\rm SOI}_{j}$ 
on optical conductivities and the magneto-optical Kerr effect is 
small~\cite{optical_conductivity_nickel_wang_callaway,moke_li_suzuki}.
At first glance, it is therefore surprising that $a_{yx}^{\rm(II)}$ is as important
as $a_{yx}^{\rm(I)}$ in Co/Pt(111) according to Fig.~\ref{fig_decompose_spincurr}.
One reason is that SOI in Pt is stronger than in Fe and Ni, 
but a second important reason is that from the 10 layers of Pt only the first few layers close to 
the interface matter for DMI while all Pt layers contribute to $a_{yx}^{\rm(I)}$ and
to $a_{yx}^{\rm(II)}$. In order to illustrate this we calculate the coefficient
$a_{yx}^{\rm (I,4)}$, which is obtained from Eq.~\eqref{eq_aj_one}
when SOI is included only in the 
3 Co layers and in the
adjacent interfacial Pt layer 
and 
artificially switched off in the other 9 Pt layers.
We plot the resulting coefficient $a_{yx}^{\rm (I,4)}$ in Fig.~\ref{fig_contribution_aI4_vs_dmi}
as a function of Fermi energy shift $\Delta E_{\rm F}$ and compare 
it to the DMI coefficient $D_{yx}$.
Fig.~\ref{fig_contribution_aI4_vs_dmi} shows that $a_{yx}^{\rm (I,4)}$ is a good approximation
for $D_{yx}$ in the region -0.75~eV$<\Delta E_{\rm F}<$0. This suggests the picture
that the essential DMI physics in transition metal bilayers is contained in $a_{yx}^{\rm(I)}$.
However, when the 5$d$ heavy metal layer is very thick, $a_{yx}^{\rm(I)}$ gets 
contaminated by $-a_{yx}^{\rm(II)}$ and therefore the correct expression in first order of SOI
is $D_{yx}^{\rm (1)}=a_{yx}^{\rm(I)}+a_{yx}^{\rm(II)}$. This
interpretation is also in line with our discussion in
Section~\ref{sec_ajII}, where we point out that $a_{yx}^{\rm(I)}$ 
and $a_{yx}^{\rm(II)}$ may individually be nonzero in nonmagnets while
their sum cancels out in nonmagnets.

\section{Summary and Outlook}
\label{sec_summary}
We show analytically that at the first order in the
perturbation by SOI DMI is given by the ground-state spin current.
As a consequence, ground-state spin currents in nonmagnetic systems
cannot exist at the first order in SOI. 
In the special case of the Rashba model they arise at the third order
in SOI. 
This clarifies the connection between the Berry-phase approach 
and the spin-current approach to DMI.
The SOI-linear contribution to DMI can be
decomposed into two contributions.
The first contribution can be understood by
mapping spin-spirals onto magnetically collinear systems by a gauge
transformation and adding spin-orbit coupling perturbatively.
We obtain an intuitive interpretation of the first contribution
as Zeeman interaction between the spin-orbit field and the 
misalignment of electron spins in magnetically noncollinear textures.
We discuss how the misalignment is related to the spin-transfer torque
and how the symmetry of DMI is related to the spin-orbit field.
Thereby, we also provide a simple explanation why DMI and the
ground-state spin current are related. 
The second contribution arises from the SOI-correction to the
velocity operator. While the SOI-correction to the velocity operator
is in principle small in transition metals, its contribution to DMI
cannot be neglected in magnetic bilayer systems with thick heavy metal layers. 
When magnetic textures are moving, the spin misalignment of electrons leads
to counter-propagating spin currents. These counter-propagating spin currents
carry energy due to their Zeeman interaction with the spin-orbit field. 
Thereby, our theory highlights the connections of DMI to spintronics
concepts such as spin-orbit fields and spin-transfer torque.
We
calculate DMI and ground-state spin currents from \textit{ab-initio} in
Mn/W(001) and Co/Pt(111) magnetic bilayers. We find that due to the strong
SOI from the heavy metal layers DMI is not well approximated by the full
ground-state spin current. 
Thereby, we illustrate the limitations of the spin-current approach to DMI
in systems with strong SOI.
However, the SOI-linear contribution to the
ground-state spin current provides a good and useful approximation for DMI
in systems with strong SOI because 
DMI is much more linear in SOI than the ground-state spin current is.

The application of electric fields or light can change the DMI 
coefficients~\cite{ultrafast_modification_exchange_interaction}.
While a complete \textit{ab-initio} theory of nonequilibrium 
exchange interactions and DMI is still missing,
an interesting application of the spin-current description of DMI is
the estimation of the variation of DMI by nonequilibrium spin currents
excited by applied electric fields or by light. 
In order to induce or to modify DMI the spin in the nonequilibrium
spin current needs to have a component perpendicular to the magnetization.
One option to generate such a spin current is 
the spin Hall effect~\cite{rmp_she}, which allows the generation of
spin currents of the order of $10^7$A/cm$^2\hbar/e$
in metals with large SOI. A spin current of this size corresponds to 
a DMI-change of the order of
0.05~meV\AA\, per atom. While this is smaller than the 
equilibrium DMI in Mn/W and Co/Pt 
by more than 2 orders of magnitude and therefore difficult to measure in 
these systems, 
DMI-changes due to the spin Hall effect
might be measurable in systems with small or zero DMI.
Using femtosecond laser pulses one can excite significantly stronger
nonequilibrium spin currents of the order 
of $10^9$A/cm$^2\hbar/e$~\cite{thz_spin_current_kampfrath}.
For such strong spin-currents the  spin-current picture of DMI leads 
to the estimate of a DMI-change of 5~meV\AA\, per atom, which 
is the order of magnitude of the equilibrium DMI in Mn/W and Co/Pt bilayer systems.
\acknowledgments
We gratefully acknowledge computing time on the supercomputers
of J\"ulich Supercomputing Center and RWTH Aachen University
as well as financial support from the programme
SPP 1538 Spin Caloric Transport
of the Deutsche Forschungsgemeinschaft.

\bibliography{spicudmi}

\begin{thebibliography}{44}
\bibitem [{not()}]{note_on_K_vs_F}%
  \BibitemOpen
  \href@noop {} {}\bibinfo {note} {Using $\partial(\beta
  g(\mathcal{E}))/\partial\beta=(\mathcal{E}-\mu)f(\mathcal{E})$ one can easily
  obtain a relation for $\partial(\beta \vn{D}_{j})/\partial\beta$ from
  Eq.~\eqref{eq_dmi_green_functions_gder}. The resulting expression for the DMI
  contribution to the grand canonical energy density $\partial(\beta
  \vn{D}_{j})/\partial\beta$ agrees to the expression reported previously by us
  in Ref.~\cite{mothedmisot} (See Eq.~(29) in
  Ref.~\cite{mothedmisot}).}\BibitemShut {Stop}%
\end{thebibliography}

\end{document}